\documentclass{article}

    \PassOptionsToPackage{numbers, compress}{natbib}
 \usepackage[position, final]{neurips_2026}
\usepackage[utf8]{inputenc} % allow utf-8 input
\usepackage[T1]{fontenc}    % use 8-bit T1 fonts
\usepackage{hyperref}       % hyperlinks
\usepackage{url}            % simple URL typesetting
\usepackage{booktabs}       % professional-quality tables
\usepackage{amsfonts}       % blackboard math symbols
\usepackage{nicefrac}       % compact symbols for 1/2, etc.
\usepackage{microtype}      % microtypography
\usepackage{xcolor}         % colors
\usepackage{subcaption}
\usepackage{hyperref}
\usepackage{url}
\usepackage{subcaption}
\usepackage{wrapfig}
\usepackage{xspace}
\usepackage{pifont}
\usepackage{booktabs}
\newcommand{\ie}{\textit{i.e.,}\xspace}
\newcommand{\eg}{\textit{e.g.,}\xspace}

\newcommand{\chatgpt}{\textsc{ChatGPT}\xspace}
\newcommand{\toolname}{\textsc{SciCodePile}}
\usepackage{makecell}
\usepackage{amsmath}
\usepackage{times}
\usepackage{enumitem}
\usepackage{latexsym}
\usepackage{graphicx}
\usepackage{subcaption}
\usepackage{tikz}

\DeclareRobustCommand{\halfcircle}{%
  \tikz[baseline=-0.6ex]{%
    \path[fill=black] (0,0) -- (90:0.8ex) arc (90:270:0.8ex) -- cycle;
    \draw (0,0) circle (0.8ex);
  }%
}

\DeclareRobustCommand{\emptycircle}{%
  \tikz[baseline=-0.6ex]{\draw (0,0) circle (0.8ex);}%
}

\DeclareRobustCommand{\fullcircle}{%
  \tikz[baseline=-0.6ex]{\fill (0,0) circle (0.8ex);}%
}

\usepackage{listings}
\usepackage{xcolor}

\title{\toolname{}: A 128GB Corpus and Executable Benchmark for Challenging Scientific Code Generation}

\author{%
  Weifeng Sun \\
  Singapore Management University\\
  \texttt{wfsun@smu.edu.sg} \\
  \And
  Ye Fan, Yuchen Chen \\
  Nanjing University \\
  \AND
  Gou Tan \\
  SUN YAT-SEN University \\
  \And
  Jieke Shi
  \thanks{Corresponding author.}\\
  Singapore Management University \\
  \And
  Yuan Yidi, Swee Liang Wong, Jonathan Pan \\
  Home Team Science \& Technology Agency \\
  \And
  David Lo \\
  Singapore Management University \\
}

\begin{document}

\maketitle

\begin{abstract}
Large language models (LLMs) excel at general-purpose code generation, yet how well they handle scientific code remains an open question. Existing datasets and benchmarks are limited in scale, domain coverage, or executable verification, leaving the true gap between current LLMs and reliable scientific code generators inadequately assessed.
To address these limitations, we present \toolname{}, the largest scientific code corpus to date, constructed from 37,737 public repositories and collectively comprising 128GB of code that spans multiple computational science disciplines. From this corpus, we further curate an executable benchmark of 200 tasks, each equipped with a sandboxed execution environment and an automated test harness for functional verification.
We evaluate 15 LLMs from both open-source and closed-source families on three tasks: prefix-to-suffix completion, fill-in-the-middle infilling, and executable code generation. Results show that scientific code generation remains highly challenging: The best CodeBLEU reaches only 38.13 and 38.37 on the two completion tasks, while the strongest model achieves just 12.30\% Pass@1 on the executable benchmark, underscoring how far current models remain from reliable scientific code generation.
To demonstrate the training utility of \toolname{}, we further show that continued pretraining on our corpus improves CodeBLEU by $\times$2.84 on scientific code completion, and instruction tuning on our data improves Pass@1 by $\times$4.79 on the executable benchmark. All code and data are available at \url{https://huggingface.co/SciCodePile}.
\end{abstract}

\section{Introduction}

Large language models (LLMs), particularly code-specialized variants such as Qwen2.5-Coder~\cite{Hui2024} and DeepSeek-Coder~\cite{Guo2024}, have substantially advanced automated programming support~\cite{Hou2024}, enabling practical applications in code completion~\cite{Wang2025} and generation~\cite{Jiang2026,Spiess2025}. However, applying these models to computational science workflows poses distinct challenges, as correctness in domains such as quantum chemistry~\cite{Niazi2025} and molecular docking~\cite{Bhat2025, Bello2025} requires not only syntactic validity but also adherence to domain-specific semantics, complex numerical logic, and project-level dependencies, which go beyond general programming competence.

\noindent\textbf{Limitations of Existing Work.}
% Recent benchmarks such as MatTools~\cite{Liu2025} and ScienceAgentBench~\cite{Chen2024} have taken important steps toward evaluating scientific code generation, but they are primarily evaluation-oriented.
Advancing LLMs for scientific code generation requires both domain-representative corpora that provide training signals for model improvement and executable benchmarks that support reliable functional evaluation.
However, existing efforts still fall short in two key respects, as summarized in Table~\ref{tab:scientific_code_dataset_compare}.
\ding{182} \textit{Limited scale or uneven domain coverage.} Most existing benchmarks are designed primarily for evaluation and remain relatively small for large-scale training (\eg SciCode contains only 80 main tasks). Although ChemPile~\cite{Mirza2025} provides a large-scale dataset spanning diverse chemical content types, its code subset is chemistry-centric and collected through keyword-based filtering over public sources, which may introduce noise and limit coverage beyond chemistry. \ding{183} \textit{Limited cross-level alignment between project intent and code artifacts.} Existing benchmarks rarely preserve explicit links between repository-level goals (\eg \texttt{README} descriptions), the functions that realize them, and the tasks derived from those functions. As a result, instruction datasets are often constructed from code content alone, overlooking project-level intent and contextual evidence, which makes it harder to obtain well-grounded and interpretable supervision.
% These limitations motivate the following question: \textbf{How far are current LLMs from reliably generating correct and executable scientific code, and do we have sufficient data to bridge this gap?}
Together, these limitations restrict both the development and the reliable assessment of LLMs for scientific code generation.

\begin{wraptable}{r}{0.55\textwidth}
\centering
\scriptsize
\vspace{-0.3cm}
\caption{\textbf{Comparison against prior scientific code datasets.}
``Pre-train'': provides a corpus suitable for pretraining or continued pretraining;
``Eval.'': includes an evaluation suite for scientific coding tasks;
``Test'': offers executable verification.
\emptycircle/\halfcircle/\fullcircle\ denote no/partial/full support.}
\label{tab:scientific_code_dataset_compare}
\setlength
\tabcolsep{0.3pt}
% \resizebox{0.50\textwidth}{!}{
\begin{tabular}{lcccccc}
\toprule
\textbf{Dataset} & \textbf{Scale} & \textbf{Domain} &
\textbf{Pre-train}  & \textbf{Eval.} & \textbf{Test} \\
\midrule
SciCode~\cite{Tian2024}      & 80 tasks      & \makecell{Multiple\\Fields}         & \emptycircle & \fullcircle & \fullcircle \\
MatTools~\cite{Liu2025}      & 49 tasks      & Materials                           & \emptycircle & \fullcircle & \halfcircle \\
ScienceAgentBench~\cite{Chen2024} & 102 tasks & \makecell{Scientific\\Discovery}    & \emptycircle & \fullcircle & \halfcircle \\
DiscoveryBench~\cite{Majumder2024} & 1,167 tasks & \makecell{Scientific\\Discovery}  & \emptycircle & \fullcircle & \emptycircle \\
MedicalCoder~\cite{Soroush2024} & 27,000+ tasks & \makecell{Medical\\Coding}      & \emptycircle & \fullcircle & \emptycircle \\
DSCodeBench~\cite{Ouyang2025} & 1,000 tasks   & \makecell{Data\\Science}            & \emptycircle & \fullcircle & \fullcircle \\
BioCoder~\cite{Tang2024}     & 2,522 tasks   & Bioinformatics                       & \halfcircle  & \fullcircle & \fullcircle \\
HPC-Coder~\cite{Nichols2024} & 1.62GB        & \makecell{HPC\\Parallel}            & \halfcircle  & \fullcircle & \fullcircle \\
ChemPile~\cite{Mirza2025}    & 15.6GB        & Chemistry                            & \fullcircle  & \emptycircle & \emptycircle \\
\midrule
\midrule
\textbf{\toolname{} (Ours)}  & \textbf{128GB}   & \textbf{Multiple Fields}         & \fullcircle & \fullcircle & \fullcircle \\
\bottomrule
\end{tabular}
\vspace{-0.2cm}
\end{wraptable}

\vspace{0.05cm}
\noindent\textbf{Our Solution.} To address this gap, we present \toolname{}, the largest scientific code corpus to date, along with an executable benchmark for computational science code generation. 
Specifically, \toolname{} incorporates three key characteristics:
\ding{182} \textit{Domain-focused repository curation at scale.} Rather than relying on keyword-based filtering alone, we design a retrieve-then-filter pipeline that combines LLM-assisted vocabulary expansion with conservative quality and relevance controls to obtain a high-precision pool of 37,737 repositories for dataset construction.
\ding{183} \textit{Multi-granularity aligned data formats.} Unlike prior datasets that construct instructions from code alone, \toolname{} provides four complementary formats, namely Code Files, README Summary, Function Instruction, and Problem–Solution Pairs, which explicitly connect repository-level intent, function-level evidence, and task-level formulations across different granularities.
\ding{184} \textit{Executable benchmark with test-driven evaluation.}
\toolname{} includes a benchmark of 200 tasks, each equipped with a sandboxed execution environment and an automated test harness. 
A model is considered successful only if its generated solution passes all tests, thereby evaluating whether the code is actually runnable and functionally correct rather than merely similar to a reference implementation.

By combining large-scale corpus construction, multi-granularity format alignment, and executable verification, \toolname{} provides both the training resources and evaluation infrastructure needed to systematically advance scientific code generation. 
This combination distinguishes our work from prior studies and offers both technical novelty and a unique perspective.

\noindent\textbf{Evaluation.} To assess the utility of the constructed corpus and benchmark, we conduct a comprehensive empirical study using 15 open-source and closed-source LLMs. The evaluation covers three realistic code-generation tasks: (i) prefix-to-suffix code completion, (ii) fill-in-the-middle infilling, and (iii) instruction-following code generation on the executable benchmark. 
For the completion tasks, we report BLEU and CodeBLEU against the reference code, while the executable benchmark is evaluated using Pass@1 and Pass@5, since correctness is determined by test execution rather than textual similarity. 
The results show that repository-level scientific code completion remains challenging for current models, with the best performance reaching 38.13 CodeBLEU in prefix completion and 38.37 CodeBLEU in fill-in-the-middle completion. 
On the executable benchmark, performance is substantially lower: The best model, \textit{GPT-5.4-mini}, achieves only 12.30\% Pass@1 and 15.50\% Pass@5, underscoring the substantial gap between syntactically plausible output and functionally correct scientific implementations.
To further validate the training utility of \toolname{}, we conduct two lightweight training experiments using small models due to resource constraints. Continued pretraining of GPT-2 (124M) on our scientific code corpus improves CodeBLEU by $\times$2.84 on code completion tasks, while instruction tuning of Qwen2.5-Coder-0.5B on F3/F4 improves Pass@1 by $\times$4.79 on the executable benchmark (1.90\% $\rightarrow$ 9.10\%), confirming that \toolname{} provides effective training signal for scientific code generation.

\vspace{0.05cm}
\noindent\textbf{Contributions.} In summary, our paper makes the following contributions: 
\begin{itemize}[leftmargin=*] 
    \vspace{-0.2cm}
    \item We construct the largest scientific code corpus to date, consisting of 37,737 public repositories organized into four complementary data formats that explicitly connect repository-level documentation with function-level evidence and task-level formulations.
    \vspace{-0.1cm}
    \item We introduce an executable benchmark of 200 tasks for instruction-following scientific code generation, each equipped with a sandboxed execution environment and an automated test harness for functional verification.
    \vspace{-0.1cm}
    \item We conduct a systematic empirical evaluation of 15 diverse LLMs on three realistic code-generation tasks, revealing that current models fall significantly short of reliable scientific code generation and providing actionable insights into their failure patterns. 
    \item We demonstrate the training utility of \toolname{} through two experiments: continued pretraining in F1 improves CodeBLEU by $\times$2.84 on scientific code completion, while instruction tuning in F3/F4 improves Pass@1 by $\times$4.79 on the executable benchmark (1.90\% $\rightarrow$ 9.10\%), confirming that \toolname{} provides an effective training signal for scientific code generation.
\end{itemize}

% We then conduct an empirical study to quantify how well current LLMs handle chemistry-domain coding under three representative tasks: (1) \emph{prefix-to-suffix} code completion, where models must continue a real file given its initial context; (2) \emph{fill-in-the-middle} completion, where a missing span must be reconstructed from surrounding context; and (3) \emph{instruction-to-code} generation grounded in real repositories. Our results show that, despite small advantages of code-specialized models, exact match remains near zero and BLEU scores stay low across tasks, highlighting a substantial gap between surface-level plausibility and repository-consistent, scientifically correct implementations.

% These findings motivate the core technical challenges that we target in subsequent milestones: providing models with richer project context (e.g., dependencies, retrieved related functions, and environment constraints), developing domain-aware correctness checks beyond text similarity (e.g., parsing/compilation success, unit tests, and numerical invariants), and ultimately building trustworthy generative assistants that can operate within heterogeneous, multi-language scientific codebases. We view the constructed corpus and the accompanying evaluations as a necessary foundation for measuring progress and enabling reliable, reproducible AI-assisted programming in computational chemistry.
\vspace{-0.4cm}
\section{Related Work}
\vspace{-0.2cm}
Recent work has explored datasets and benchmarks for scientific programming and AI-assisted research.
SciCode~\cite{Tian2024} introduces a scientist-curated benchmark with 80 research coding problems across multiple domains, each accompanied by executable tests.
MatTools~\cite{Liu2025}, ScienceAgentBench~\cite{Chen2024}, and DiscoveryBench~\cite{Majumder2024} evaluate LLMs in scientific discovery workflows, including tool usage and research-oriented problem solving.
Other benchmarks focus on specific domains, such as MedicalCoder~\cite{Soroush2024} for medical coding, DSCodeBench~\cite{Ouyang2025} for data science programming, BioCoder~\cite{Tang2024} for bioinformatics code generation, and HPC-Coder~\cite{Nichols2024} for high-performance computing.
Another line of work provides large-scale scientific corpora for model pretraining.
For example, ChemPile~\cite{Mirza2025} collects a large-scale chemistry dataset spanning scientific text, molecular representations, and code artifacts.
However, existing resources typically emphasize either evaluation benchmarks (e.g., SciCode, MatTools) or domain-specific pretraining corpora (e.g., ChemPile), but rarely provide both large-scale repository-grounded data and executable test-based evaluation.
Our work addresses this gap by constructing a multi-format scientific code corpus together with an executable benchmark derived from real-world repositories.

\begin{figure*}[t]
   \centering
   \includegraphics[width=0.98\textwidth]{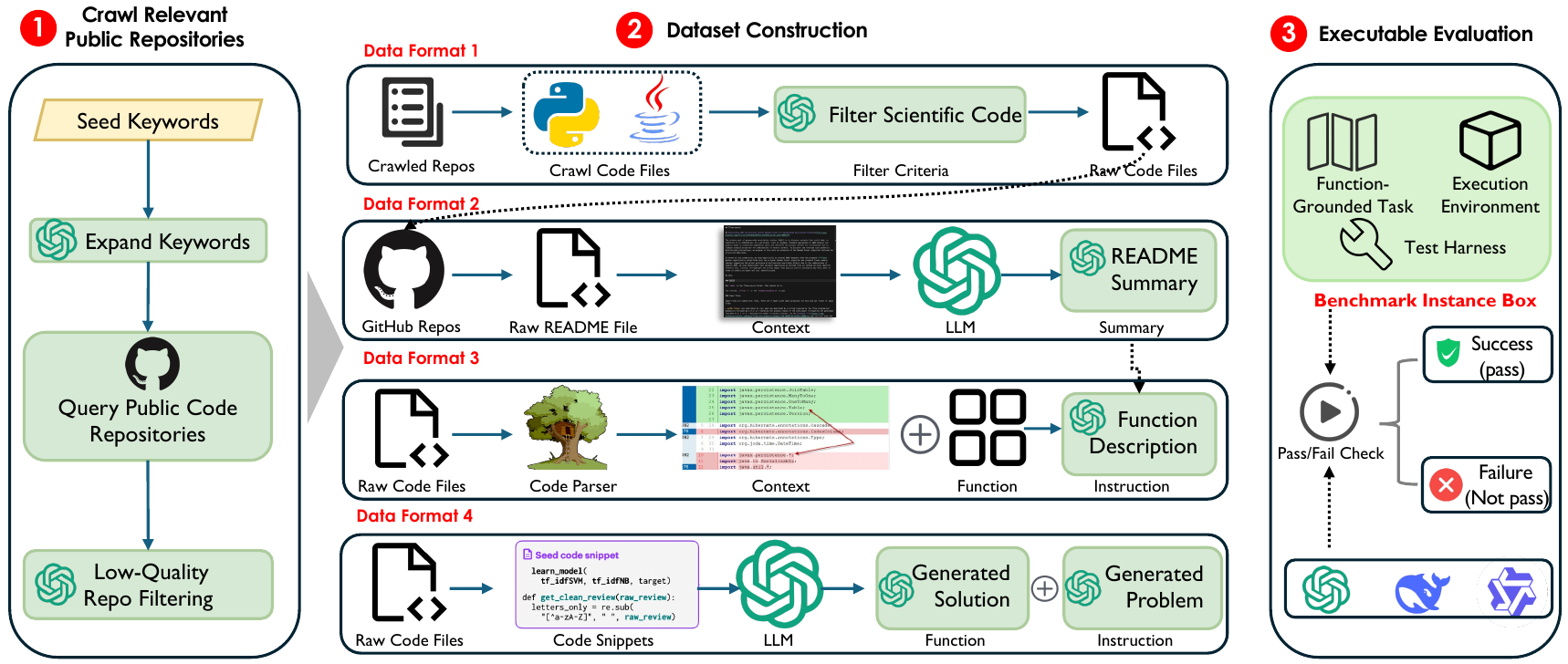}
   \caption{Overall workflow of dataset construction.}
   \label{fig:overall}
\vspace{-0.5cm}
\end{figure*}

\vspace{-0.3cm}
\section{\toolname{}}
\label{sec:method}
\vspace{-0.3cm}
\noindent
This section presents the end-to-end framework used to construct \toolname{}, a multi-format corpus and benchmark for computational science code generation. As shown in Figure~\ref{fig:overall}, the framework is organized into three tightly-coupled components: (i) \emph{crawling relevant public repositories} to build a broad candidate pool; (ii) \emph{deriving four aligned data formats} from the curated repositories to support pretraining and instruction-oriented supervision at multiple granularities; and (iii) \emph{constructing an executable benchmark} to evaluate whether generated code satisfies functional requirements by passing tests in sandboxed execution environments.
We use a fixed set of models for different roles in the construction pipeline; the model selections and justifications are provided in Appendix~\ref{sec:appendix_model_roles}.

\vspace{-0.3cm}
\subsection{Component 1: Crawling Relevant Public Repositories}
\label{sec:component1}
Component~1 aims to construct a broad yet domain-focused repository pool for computational science code, maximizing coverage while maintaining relevance.
Given the noise and heterogeneity of public repositories, we formulate the process as \emph{retrieve-then-filter}: we first retrieve candidate repositories using an expanded scientific vocabulary, then apply quality and relevance controls to obtain a clean input set for \textbf{Component~2}.

\vspace{-0.3cm}
\paragraph{Seed keyword design.}
We initialize repository retrieval with 198 seed keywords curated by the authors based on established computational science taxonomies (the full list is provided in Appendix~\ref{sec:keyword_list}), spanning major science subfields and their associated tool ecosystems, including: (i) chemistry-focused topics (\eg quantum chemistry, molecular dynamics); (ii) life science and bioinformatics workflows (\eg genomics, proteomics); (iii) pharmaceutics and biomedical modeling (\eg pharmacokinetics); (iv) computational and numerical methods (\eg finite element method); and (v) AI/ML-driven scientific modeling (\eg generative models). This design balances precision and coverage for large-scale repository retrieval.

\vspace{-0.3cm}
\paragraph{LLM-assisted vocabulary expansion.}
To improve coverage of scientific terminology, we use \chatgpt-4o to expand the 198 seed keywords into a richer set of 213 queries by generating three types of variants: (i) synonyms and spelling alternatives; (ii) abbreviations and their full-form mappings; and (iii) closely related subtopics and domain-specific terms. The prompt used for keyword expansion is provided in Appendix~\ref{sec:append_f1}.

\vspace{-0.3cm}
\paragraph{Repository retrieval and metadata.}
Using the expanded keyword set, we query GitHub\footnote{\url{https://docs.github.com/en/rest/search}} via its search APIs to retrieve candidate repositories. The crawl covers repositories available on GitHub as of December 2025 and yields 1,311,568 candidates. For each repository, we collect metadata for subsequent filtering and analysis, including the URL, description, topics (when available), primary languages, repository size, and star and fork counts.

\vspace{-0.3cm}
\paragraph{Quality and relevance filtering.}
We apply conservative heuristics to remove repositories unlikely to contain substantive scientific implementations. Specifically, (i) we exclude repositories with fewer than 10 stars, following prior work~\cite{Sun2023, Wen2019}, to filter low-quality projects; (ii) we remove documentation-only repositories dominated by non-source content; and (iii) to enforce scientific relevance beyond keyword matches, we prompt Qwen3-32B~\cite{Yang2025} to judge whether the \texttt{README} describes a repository that actually implements computational-science methods or workflows, rather than merely mentioning related keywords, and discard those predicted as non-relevant. The prompt used for this relevance check is provided in Appendix~\ref{sec:append_f1}. After filtering, we retain 37,737 repositories covering 178 scientific keywords for Component~2.

\vspace{-0.3cm}
\subsection{Component 2: Deriving Four Aligned Data Formats}
\label{sec:component2}
\noindent
Component~2 transforms the curated repositories from Component~1 into four aligned data formats that support (i) domain-adaptive pretraining, (ii) instruction-oriented supervision, and (iii) downstream benchmark construction. We first crawl and normalize source files to build a raw code corpus (F1), then derive three progressively more structured formats (F2–F4) that capture repository-level intent, function-level semantics, and snippet-level instruction–solution signals.

\vspace{-0.3cm}
\paragraph{F1: Raw Code Files.}
\label{sec:format1}
\textbf{\textit{Motivation.}}
F1 provides a cleaned collection of scientific code files designed for continued pretraining and corpus analysis.
\textbf{\textit{Method.}} For each repository snapshot, we retain files using conservative heuristics: (i) a language whitelist covering major scientific programming languages (\eg Python, C/C++, Fortran, and Julia) as well as widely used scripting and configuration formats; and (ii) path-based exclusion rules to remove non-source or auxiliary content (\eg \texttt{.git/}, \texttt{build/}, \texttt{dist/}, and \texttt{third\_party/}). We further normalize whitespace, compute content hashes, and remove near-duplicate files using SimHash~\cite{Charikar2002} to reduce redundancy introduced by forks, templates, and repeated headers. The resulting F1 corpus contains 125GB of code spanning approximately 3.26M files.

\vspace{-0.3cm}
\paragraph{F2: \texttt{README} Summary.}
\label{sec:format2}
\textbf{\textit{Motivation.}}
F2 distills repository-level context, including project intent, dependencies, and I/O conventions, into structured representations that provide project-level grounding for downstream formats, particularly F3 function descriptions.
\textbf{\textit{Method.}}
We extract each repository's \texttt{README} file and truncate long documents to the first 8,000 tokens. We then prompt Qwen3-32B to produce a structured summary covering: (i) project goals and task category; (ii) key methods or algorithms; (iii) dependencies and major libraries; (iv) input/output artifacts; and (v) the expected runtime environment. The \texttt{README}-summary prompt is provided in Appendix~\ref{sec:append_f2}.

\vspace{-0.3cm}
\paragraph{F3: Function Instruction.}
\label{sec:format3}
\textbf{\textit{Motivation.}}
F3 pairs real functions with evidence-constrained natural-language descriptions, supporting function retrieval, documentation generation, and instruction tuning under domain constraints.
\textbf{\textit{Method.}}
We extract functions from F1 using \texttt{tree-sitter}~\cite{treesitter} across multiple languages and retain traceability metadata, including the repository, file path, span, function name or qualified name, and raw code. To prioritize domain-relevant scientific implementations, we score each function using two embedding-based signals computed with Qwen3-embedding: $md_{score}$, the cosine similarity between the domain keyword set and the repository \texttt{README}; and $code_{score}$, the similarity between the same keyword set and the function body. Candidates are ranked by
\begin{equation}
    Final_{score} = md_{score} \times code_{score}
\end{equation}
For each selected function, we assemble lightweight context from the function code, extracted dependencies (\eg imports and salient API calls), and the corresponding \texttt{README} summary from F2. 
We then prompt \chatgpt-5 to generate a docstring-style description grounded in this evidence.
To improve precision, we use NatureLM~\citep{Xia2025}, a scientific domain model, as an automatic verifier to assess whether the generated description is both (i) consistent with the provided context, and (ii) relevant to the target scientific domains.
Generations that fail these checks are treated as low-confidence and filtered out.
The function-instruction generation prompt is provided in Appendix~\ref{sec:appd_f3}, and an example F3 instance is shown in Appendix~\ref{sec:format3-example}.

\vspace{-0.3cm}
\paragraph{F4: Problem–Solution Pairs.}
\label{sec:format4}
\textbf{\textit{Motivation.}}
F4 constructs function-grounded problem–solution pairs using real code as anchors, leveraging open-source references to elicit diverse yet controllable instruction-style training signals.
\textbf{\textit{Method.}} Following the OSS-Instruct principle~\citep{Wei2023}, we sample 20,000 domain-relevant functions from F3 and randomly extract code snippets within each function to cover diverse implementation regions. Given each snippet and its repository context (imports, dependencies, and \texttt{README} summary from F2), we prompt \chatgpt-5 to infer a plausible task specification and generate the corresponding solution, explicitly disallowing dependencies absent from the provided context. To further increase diversity, we apply data augmentation strategies including instruction paraphrasing~\cite{Abaskohi2023}, alternative coding styles~\cite{Munson2024}, and multi-sample decoding~\cite{Wang2023}. The prompt used for F4 construction is provided in Appendix~\ref{sec:appd_f4}.

\vspace{-0.3cm}
\subsection{Executable Benchmark Construction}
\label{sec:benchmark}

\paragraph{Task Definition.}
We construct an executable benchmark for evaluating large language models on scientific code generation, specifically the task of translating natural-language scientific requirements into runnable code implementations under domain constraints. Each instance provides an automatic pass/fail signal in a sandboxed environment.

\vspace{-0.3cm}
\paragraph{From Repository Functions to Candidate Tasks.} We construct candidate tasks from F3, which provides function-level code–description alignments grounded in real-world repositories. To reduce environmental dependencies and ensure self-contained evaluation, we focus on Python functions and restrict candidates to pure functions that do not depend on implicit class context. Candidates are ranked by $Final_{score}$, and the top-ranked subset is converted into HumanEval-style~\cite{Chen2021} task prompts, each consisting of a function signature and its natural-language specification (as shown in Appendix~\ref{sec:appd_f5}).

\vspace{-0.3cm}
\paragraph{Automatic Test and Environment Synthesis.}
To make tasks executable, we automatically synthesize a minimal execution setup and corresponding test harness for each candidate. Conditioned on the function specification and reference implementation, \chatgpt-5.3-codex is prompted to generate test assertions over the target function. To ensure sandbox compatibility, external dependencies are replaced with lightweight stubs, and tests are confined to standard-library functionality.

\vspace{-0.3cm}
\paragraph{Runnable Verification.}
All tasks are executed under strict isolation using a subprocess-based runner with timeout control. For each instance, we run the reference implementation, synthesized setup, and test suite end-to-end, retaining only those that complete successfully without runtime errors. After filtering, the benchmark comprises 200 fully executable instances. To further ensure quality, we manually review all retained instances to verify that the task description, setup, and tests are coherent and faithful to the underlying scientific code. An example instance is provided in Appendix~\ref{sec:execute_example}, and a manual audit of test case validity is reported in Appendix~\ref{sec:app_test_audit}.

We acknowledge two design trade-offs in the current executable benchmark. First, restricting to pure Python functions with stub-based dependencies simplifies evaluation relative to real scientific workflows; however, this is a necessary condition for reliable automated evaluation at scale, consistent with HumanEval~\cite{Chen2021} and SciCode~\cite{Tian2024}.
Second, the benchmark comprises 200 tasks due to resource constraints in human validation and sandbox construction; scaling remains an important direction for future work.

\begin{figure}[t]
    \centering
    \begin{subfigure}[t]{0.50\textwidth}
        \centering
        \includegraphics[width=\textwidth]{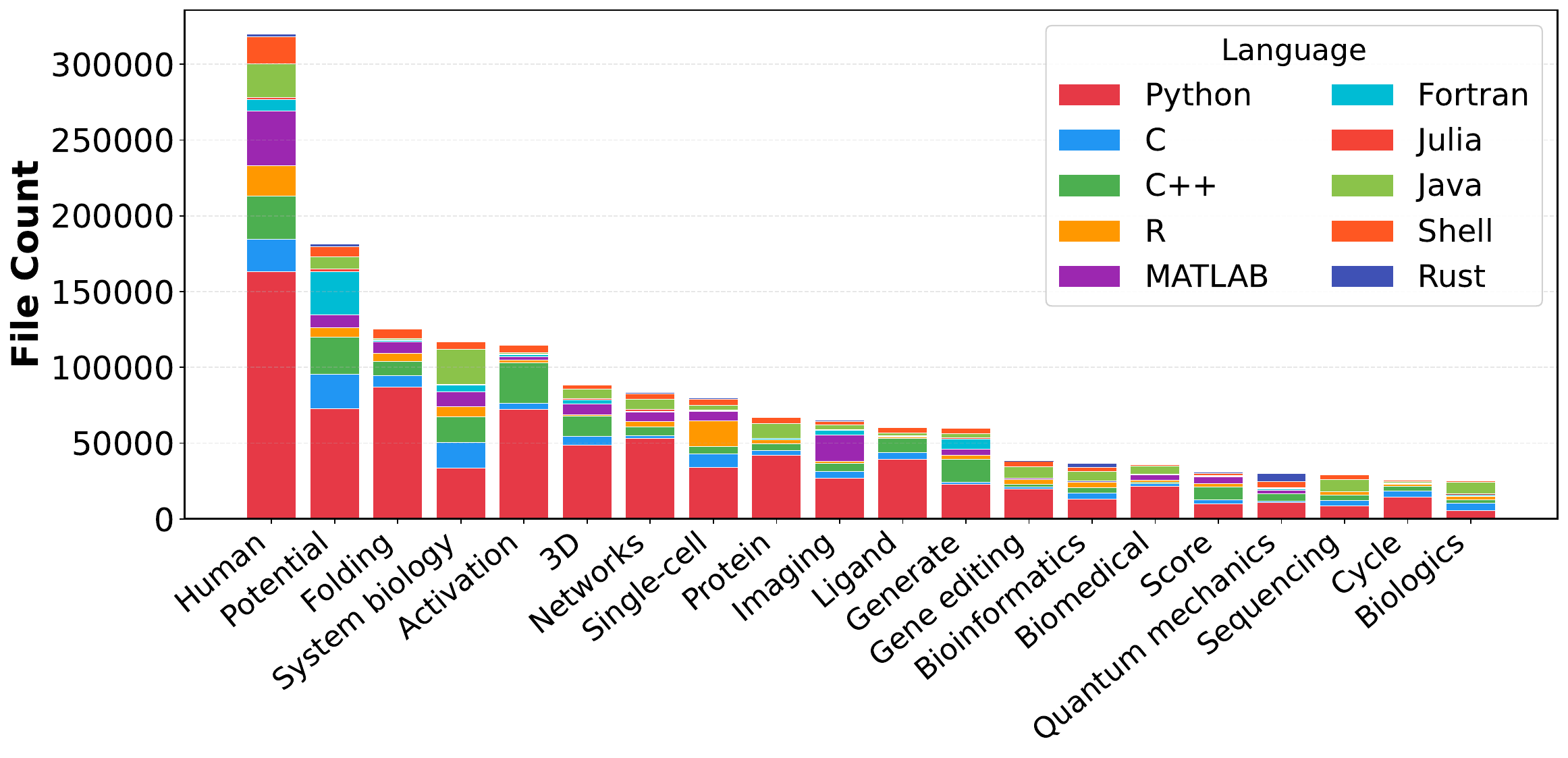}
        \caption{File-level distribution across the top-20 scientific domains with programming-language composition.}
        \label{fig:domain_language_distribution}
    \end{subfigure}
    \hfill
    \begin{subfigure}[t]{0.40\textwidth}
        \centering
        \includegraphics[width=\textwidth]{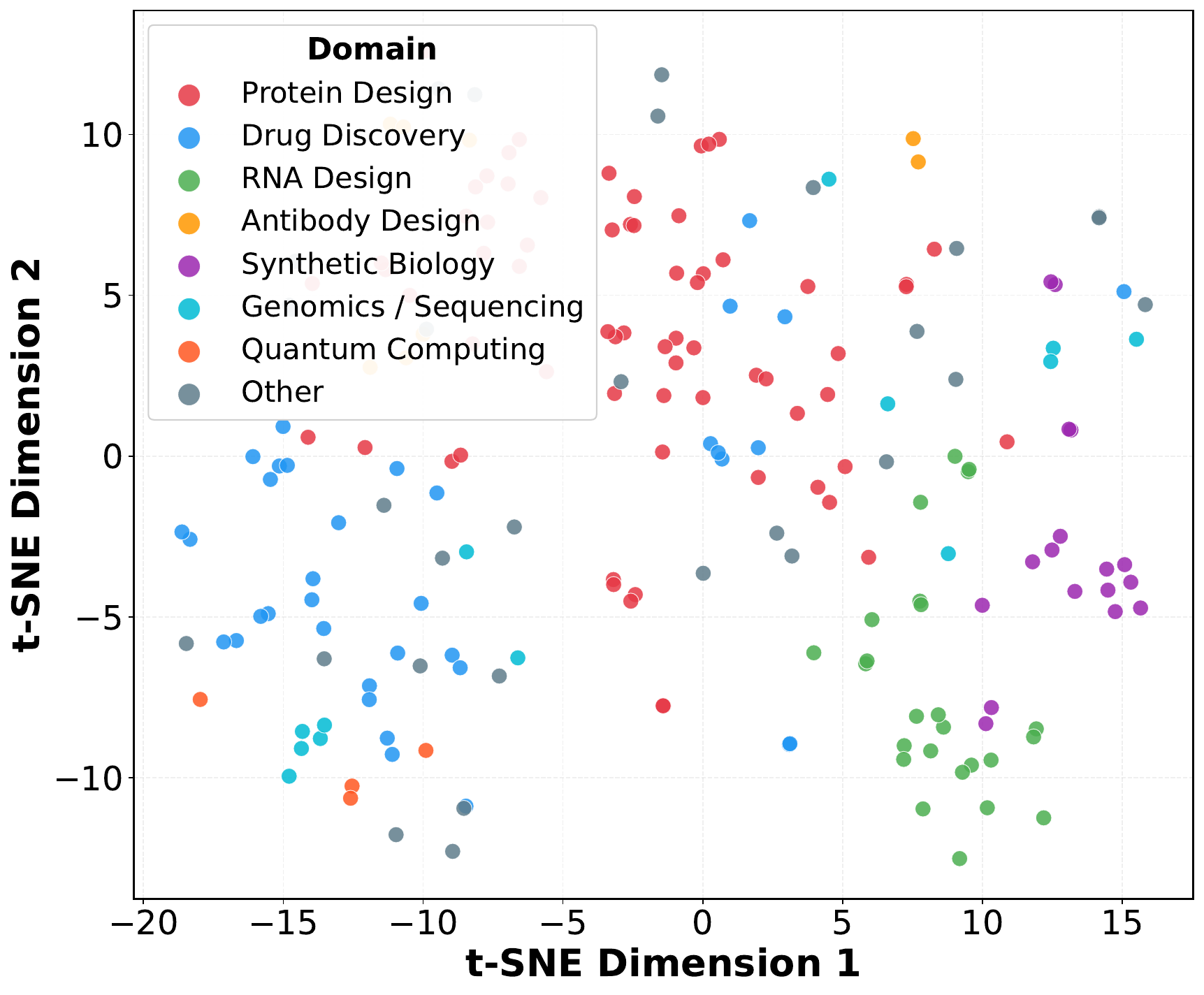}
        \caption{t-SNE visualization of prompt embeddings in the executable benchmark.}
        \label{fig:benchmark_tsne}
    \end{subfigure}
    \caption{Overview of domain-language distribution and benchmark prompt embedding space.}
    \label{fig:domain_and_benchmark}
    \vspace{-0.5cm}
\end{figure}

\vspace{-0.3cm}
\subsection{Dataset Summary}
\label{sec:dataset_stats}

We release a multi-format scientific code resource comprising four complementary data formats. \textbf{F1} contains 125GB of deduplicated raw code collected from 37,737 domain-relevant repositories, providing a large-scale corpus for continued pretraining. To provide project-level context, we additionally construct 37,737 repository-level \texttt{README} summaries, \ie \textbf{F2}, using standardized fields (\eg project overview and main features) to support downstream analysis. \textbf{F3} contains 500,000 function-level instances (2.84GB in total), derived via multi-language parsing and relevance assessment, enabling instruction-following at the function granularity. \textbf{F4} further provides 20,000 instruction–solution pairs (496MB in total), designed to increase the diversity of training signals while remaining grounded in real repository code. Our \textbf{executable benchmark} contains 200 instances derived from real-world repositories, each equipped with an execution setup and test harness. On average, each test suite contains 7.3 assertions, ensuring that candidate implementations are evaluated beyond trivial cases.
The model assignments for each pipeline stage and the corresponding stage-wise quality control mechanisms are detailed in Appendix~\ref{sec:appendix_model_roles} and Appendix~\ref{sec:appendix_quality}, respectively.

\vspace{-0.3cm}
\section{Dataset Statistics and Analysis}
\label{sec:statistics}

% In this section, we present statistical analyses of the constructed scientific code corpus to characterize its domain coverage and programming-language distribution.
\vspace{-0.3cm}
\paragraph{Domain and programming language distribution.}
Figure~\ref{fig:domain_language_distribution} summarizes the file-level distribution across the top-20 scientific domains along with their language composition. Python dominates nearly all domains, underscoring its central role in scientific computing workflows. C/C++ account for a larger share in domains that demand performance-critical numerical kernels (\eg \textit{Folding} and \textit{Activation}), while R and MATLAB are more prevalent in data-centric areas such as \textit{System Biology} and \textit{Bioinformatics}. Overall, the corpus spans diverse scientific domains and programming paradigms, reflecting the realistic heterogeneity of open-source scientific software.

\vspace{-0.3cm}
\paragraph{Semantic diversity of executable benchmark tasks.}
To examine the semantic coverage of the executable benchmark, we encode each task prompt using the all-mpnet-base-v2 embedding model\footnote{\url{https://huggingface.co/sentence-transformers/all-mpnet-base-v2}} and project the resulting embeddings into two dimensions via t-SNE~\cite{VanderMaaten2008}. As shown in Figure~\ref{fig:benchmark_tsne}, the prompts spread broadly across the embedding space rather than collapsing into a single cluster. Notably, domains such as protein design, drug discovery, RNA design, and genomics occupy distinct regions, indicating substantial variation in both task semantics and requirements. This distribution confirms that our benchmark spans a diverse spectrum of real-world scientific programming scenarios.

\begin{figure}[t]
  \centering
  \begin{subfigure}[b]{0.45\linewidth}
    \centering
    \includegraphics[width=\linewidth]{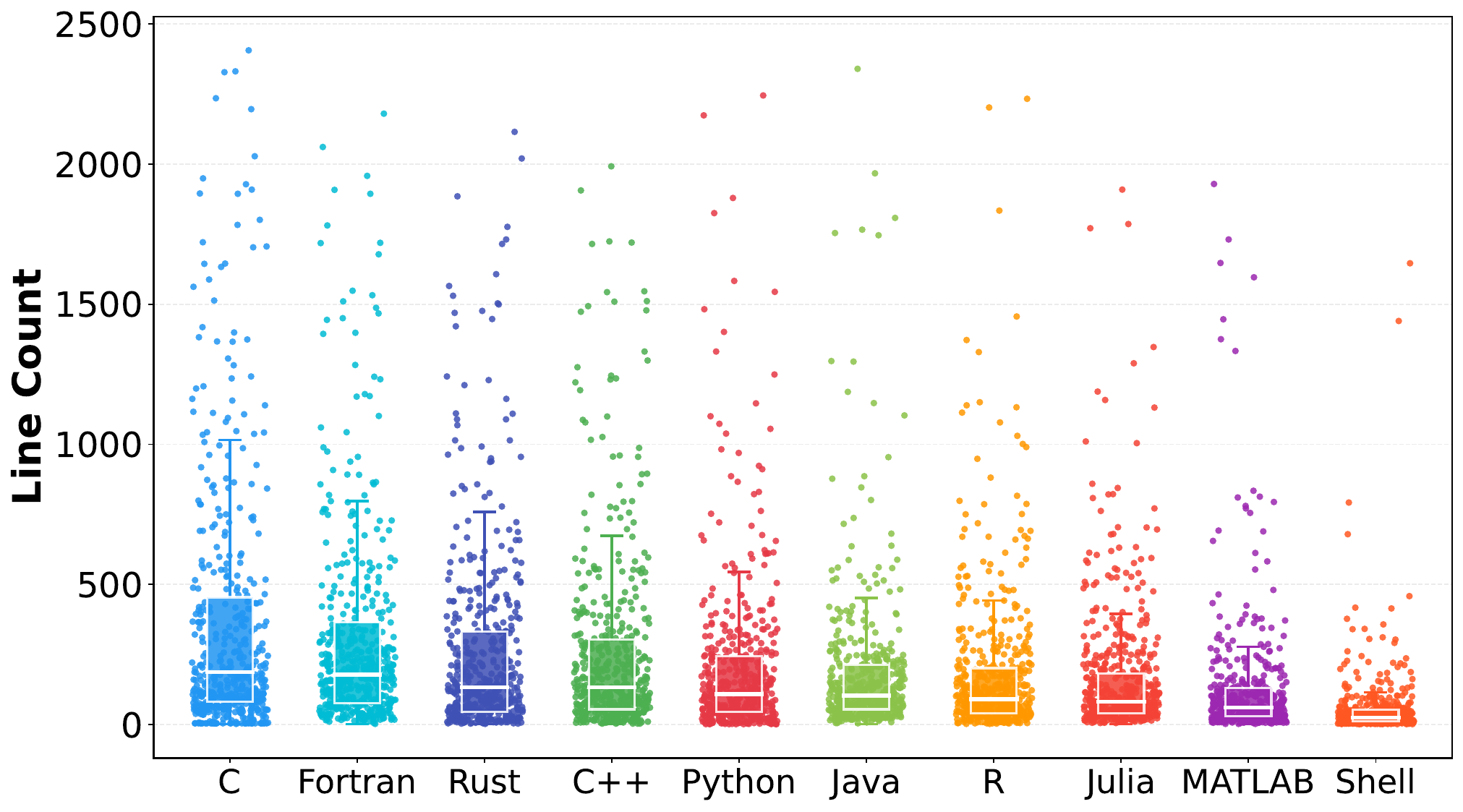}
    \caption{Line-count distribution of source files.}
    \label{fig:language_statistics_a}
  \end{subfigure}
  \hfill
  \begin{subfigure}[b]{0.45\linewidth}
    \centering
    \includegraphics[width=\linewidth]{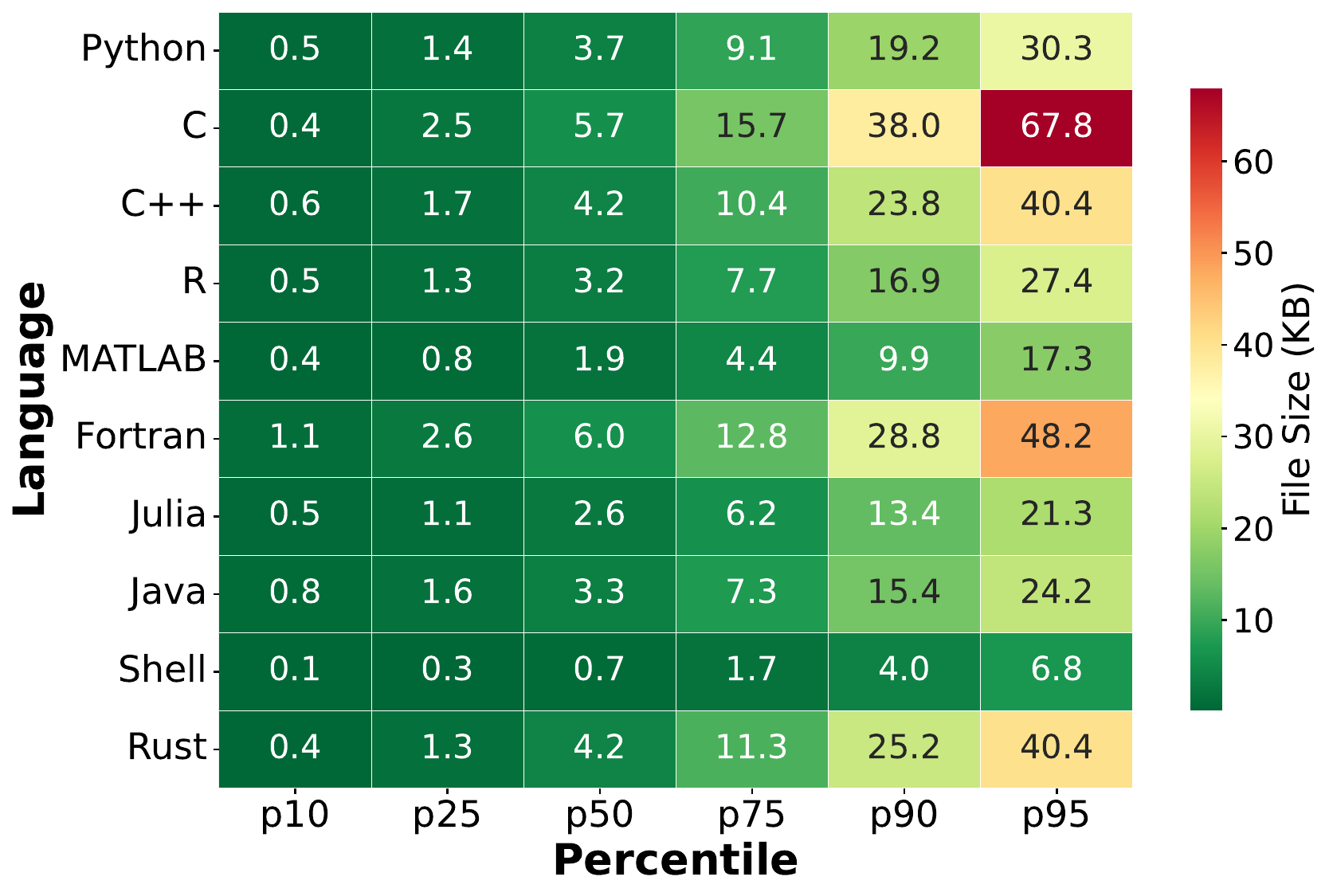}
    \caption{File-size percentiles (KB) across languages.}
    \label{fig:language_statistics_b}
  \end{subfigure}
    \caption{Programming-language characteristics in the collected scientific code corpus: (a) distribution of file line counts, and (b) percentile statistics of file sizes.}
  \label{fig:language_statistics}
  \vspace{-0.6cm}
\end{figure}

\vspace{-0.3cm}
\paragraph{File length and size characteristics.}
Figure~\ref{fig:language_statistics} further analyzes file-level characteristics across programming languages. Figure~\ref{fig:language_statistics_a} shows the distribution of source-code line counts. Languages such as \textit{C} and \textit{Fortran} exhibit higher median line counts (around 200 lines) and broader distributions. In contrast, scripting-oriented languages such as \textit{Shell} and \textit{MATLAB} tend to contain much shorter files, with most instances concentrated below 50 lines. Across all languages, the distributions display a clear long-tail pattern. Figure~\ref{fig:language_statistics_b} reports file-size percentiles computed from the full 3.2M files. \textit{C} reaches the largest upper range, with a file size of 67.8KB at the 95th percentile, followed by \textit{Fortran} at 48.2KB.
In contrast, scripting languages remain significantly smaller, with \textit{Shell} files reaching only 6.8KB at the 95th percentile.
Overall, the median file size (i.e., the 50th percentile) of most languages falls between 2KB and 6KB, suggesting that typical scientific code files remain relatively compact despite the presence of large outliers.

% \begin{figure}[h]
%     \centering
%     \includegraphics[width=0.38\textwidth]{graph/tsne_benchmark.pdf}
%     \vspace{-0.6mm}
%     \caption{t-SNE visualization of prompt embeddings in the executable benchmark.}
%     \label{fig:benchmark_tsne}
%     \vspace{-0.5cm}
% \end{figure}

\vspace{-0.2cm}
\paragraph{Semantic distinction from general coding benchmarks.}
To characterize the scientific specificity of our benchmark, we compare its prompt characteristics against HumanEval~\cite{chen2021evaluating}. The median prompt length in \toolname{} is approximately 304 words, compared to 51 words in HumanEval. Applying a scientific terminology lexicon covering key terms in biology, chemistry, and molecular modeling, 197 out of 200 tasks (98.5\%) in our benchmark contain at least one scientific term, compared to only 6 out of 164 tasks (3.7\%) in HumanEval. The most frequent terms include \textit{protein} (103 occurrences) and \textit{pdb} (65), while HumanEval contains virtually none. These results confirm that our benchmark captures domain-specific challenges that general coding benchmarks do not assess.

\vspace{-0.4cm}

\section{Experiments}
\label{sec:experiment}

% \subsection{Research Questions}

% We conduct an empirical study to evaluate how well existing code LLMs perform on scientific code generation tasks derived from real-world repositories. Our study is guided by three research questions:
% \begin{itemize}[leftmargin=*]
%     \vspace{-0.2cm}
%     \item \textbf{RQ1 (Prefix-to-Suffix).} How accurately can LLMs generate the code suffix given a prefix extracted from our crawled scientific code corpus?
%     \vspace{-0.6cm}
%     \item \textbf{RQ2 (Fill-in-the-Middle).} How accurately can LLMs recover missing middle segments of scientific code files from our crawled dataset?
%     \vspace{-0.2cm}
%     \item \textbf{RQ3 (Executable Evaluation).} How effectively do LLMs generate runnable scientific code that passes executable tests in our benchmark?
% \end{itemize}

\vspace{-0.3cm}
\subsection{Experimental Setup}
\label{sec:expe_setup}

\vspace{-0.2cm}
We evaluate models on prefix-to-suffix (\textbf{PTS}) and fill-in-the-middle (\textbf{FIM}) code completion tasks derived from our scientific code corpus. We randomly sample 1,000 class-level code files from the collected corpus F1 to construct evaluation instances.
For \textbf{PTS}, we use the first 1,024 tokens of each file as the input prefix and require the model to generate the subsequent 256 tokens as the target suffix. For \textbf{FIM}, we mask a 256-token middle span and provide the surrounding 512-token prefix and suffix as context, requiring the model to recover the missing segment. All models are evaluated in a zero-shot setting without task-specific fine-tuning.

We further evaluate models on the \textbf{executable benchmark} constructed in Section~\ref{sec:benchmark}. For each task, the model generates candidate solutions that are executed in a sandboxed environment alongside the corresponding setup code and test suite. 
% A solution is deemed correct only if it passes all test cases. 
% All experiments are conducted under a unified evaluation framework to ensure consistent execution and fair comparison across models.

\vspace{-0.3cm}
\subsection{Evaluation Metrics}

% We adopt different metrics for different settings.

\noindent\textbf{PTS and FIM Settings.}
We report \textbf{BLEU}~\cite{Papineni2002} and \textbf{CodeBLEU}~\cite{ren2020codebleu}.
BLEU measures the $n$-gram overlap between the generated code and the reference.
CodeBLEU is a code-specific evaluation metric that extends BLEU by incorporating weighted $n$-gram match, abstract syntax tree (AST) match, and data-flow match, capturing both surface-level similarity and semantic properties of code.

\noindent\textbf{Executable Benchmark.}
We report \textbf{Pass@1} and \textbf{Pass@5}, following the HumanEval protocol~\cite{Chen2021}.
Given $n$ generated solutions and $c$ correct solutions that pass the test cases, Pass@$k$ is estimated as
\vspace{-0.1cm}
\begin{equation}
\mathrm{Pass@}k = 1 - \frac{\binom{n-c}{k}}{\binom{n}{k}}
\end{equation}
% Pass@1 evaluates single-shot correctness, while Pass@5 measures the probability that at least one of the top-$k$ generated solutions passes all tests.

\vspace{-0.2cm}
\subsection{Target Models}
\label{sec:target_models}

We evaluate 15 models spanning six model families, including both general-purpose LLMs and code-specialized models.
Our open-source baselines include DeepSeek-R1, DeepSeek-R1-Distill-Qwen-1.5B~\cite{Guo2025}, Qwen3-14B-Think~\cite{Yang2025}, DeepSeek-Coder-1.3B/6.7B~\cite{Guo2024}, StarCoder2-3B/7B/15B~\cite{Lozhkov2024}, Qwen2.5-Coder-7B-Instruct~\cite{Hui2024}, and CodeLlama-7B/13B~\cite{Roziere2023}.
We further include closed-source API models, including \chatgpt-4o, o3-mini, \chatgpt-5, and \chatgpt-5.4-mini.

\vspace{-0.3cm}
\subsection{Experimental Results}
\label{sec:expe_results}
% \subsubsection{Answer to RQ1}

% \input{tables/rq2}

\begin{table}[t]
    \centering
    \scriptsize
    \setlength{\tabcolsep}{1.5pt}
    \renewcommand{\arraystretch}{0.92}
    \caption{PTS and FIM results on the scientific code corpus.}
    \label{tab:pts_fim}
    \vspace{-2mm}
    \begin{subtable}[t]{0.49\textwidth}
        \centering
        \caption{PTS results.}
        \vspace{-0.1cm}
        \label{tab:rq2}
        \begin{tabular}{lcc}
            \toprule
            \textbf{LLM} & \textbf{BLEU} & \textbf{CodeBLEU} \\
            \midrule
            DeepSeek-R1-Distill-Qwen-1.5B & 0.96 & 17.35 \\
            Qwen3-14B (Think) & 14.84 & 28.06 \\
            GPT-4o & 17.65 & 31.39 \\
            GPT-5 & 15.80 & 29.27 \\
            o3-mini & 12.53 & 27.26 \\
            DeepSeek-Coder-1.3B & 18.14 & 32.42 \\
            DeepSeek-Coder-6.7B & 22.49 & 36.44 \\
            StarCoder2-3B & 20.69 & 34.00 \\
            StarCoder2-7B & 21.34 & 34.98 \\
            StarCoder2-15B & 23.44 & 37.05 \\
            Qwen2.5-Coder-7B-Instruct & 25.24 & 38.13 \\
            CodeLlama-7B & 21.12 & 35.18 \\
            CodeLlama-13B & 22.68 & 36.49 \\
            \bottomrule
        \end{tabular}
    \end{subtable}
    \hfill
    \begin{subtable}[t]{0.49\textwidth}
        \centering
        \caption{FIM results.}
        \vspace{-0.1cm}
        \label{tab:rq3}
        \begin{tabular}{lcc}
            \toprule
            \textbf{LLM} & \textbf{BLEU} & \textbf{CodeBLEU} \\
            \midrule
            DeepSeek-R1-Distill-Qwen-1.5B & 1.06 & 16.47 \\
            Qwen3-14B (Think) & 9.72 & 22.83 \\
            GPT-4o & 14.68 & 27.31 \\
            GPT-5 & 17.05 & 30.93 \\
            o3-mini & 9.80 & 23.80 \\
            DeepSeek-Coder-1.3B & 3.41 & 15.89 \\
            DeepSeek-Coder-6.7B & 3.24 & 15.35 \\
            StarCoder2-3B & 20.36 & 34.04 \\
            StarCoder2-7B & 21.40 & 35.02 \\
            StarCoder2-15B & 23.56 & 37.05 \\
            Qwen2.5-Coder-7B-Instruct & 26.39 & 38.37 \\
            CodeLlama-7B & 3.60 & 15.60 \\
            CodeLlama-13B & 3.91 & 15.65 \\
            \bottomrule
        \end{tabular}
    \end{subtable}
    \vspace{-7mm}
\end{table}

\vspace{-0.3cm}
\textbf{\ding{182} Prefix-to-Suffix Setting.}
Table~\ref{tab:rq2} reports the prefix-to-suffix code completion results on the scientific code corpus using BLEU and CodeBLEU.
Overall, the results show that scientific code completion from prefix-only context remains challenging, since models must continue repository-level logic from incomplete context and need to infer project conventions, external dependencies, and domain-specific computation patterns that are not fully exposed in the prefix.
Among all models, \textit{Qwen2.5-Coder-7B-Instruct} achieves the best overall performance, with the highest BLEU (25.24) and CodeBLEU (38.13).
It is closely followed by \textit{StarCoder2-15B}.
This suggests that code-specialized models are generally better at modeling sequential code continuation in scientific repositories than general-purpose LLMs.

A clear scaling trend can also be observed within several model families.
For example, \textit{DeepSeek-Coder} improves from 18.14/32.42 to 22.49/36.44 when scaling from 1.3B to 6.7B, and \textit{StarCoder2} improves steadily from 34.00 to 37.05 in CodeBLEU as model size increases from 3B to 15B.
These trends indicate that larger code models are better able to capture the long-range regularities and structural patterns required for continuation-style scientific code generation.
Although strong closed-source models such as \textit{GPT-4o} and \textit{GPT-5} achieve competitive scores, they still lag behind the strongest code-specialized models.
% Overall, prefix-to-suffix completion appears to favor models with stronger sequential code modeling ability, but even the best systems remain far from reliably reproducing repository-consistent scientific code.

% \subsubsection{Answer to RQ2}

% \input{tables/rq3}

\begin{wraptable}{r}{0.45\textwidth}
    \centering
    \scriptsize
    \vspace{-0.4cm}
    \caption{Generation results on executable benchmark.}
    % \vspace{-2mm}
    \label{tab:rq4}
    \begin{tabular}{lcc}
        \toprule

        \textbf{LLM} & \textbf{Pass@1} & \textbf{Pass@5} \\

        \midrule

        DeepSeek-R1-Distill-Qwen-1.5B & 2.70\% & 7.00\% \\
        
        Qwen3-14B (Think) & 8.30\% & 10.50\% \\

        DeepSeek-Coder-1.3B & 2.80\% & 7.50\% \\

        DeepSeek-Coder-6.7B & 6.10\% & 10.00\% \\

        StarCoder2-3B & 0.50\% & 1.5\% \\

        StarCoder2-7B & 0.30\% & 1.00\% \\

        StarCoder2-15B & 0.30\% & 1.50\% \\

        Qwen2.5-Coder-7B-Instruct & 6.10\% & 8.50\% \\

        CodeLlama-7B & 7.20\% & 10.50\% \\

        CodeLlama-13B & 6.60\% & 11.00\% \\

        GPT-4o & 7.50\% & 10.00\% \\

        GPT-5 & 5.70\% & 8.50\% \\

        o3-mini & 10.50\% & 12.00\% \\

        DeepSeek-R1 & 8.50\% & 12.50\% \\

        GPT-5.4-mini & 12.30\% & 15.50\% \\

        \bottomrule
    \end{tabular}
    \vspace{-5mm}
\end{wraptable}

\noindent \textbf{\ding{183} Fill-in-the-Middle Setting.}
Table~\ref{tab:rq3} reports the fill-in-the-middle (FIM) code completion results.
Overall, the task remains difficult, and the model ranking differs from that in prefix completion: the availability of both left and right context substantially benefits some models, while others still struggle to reconstruct the missing span.
\textit{Qwen2.5-Coder-7B-Instruct} again achieves the best overall performance, with the highest BLEU (26.39) and CodeBLEU (38.37).
However, these gains are not uniform across model families.
In particular, \textit{DeepSeek-Coder-1.3B/6.7B} and \textit{CodeLlama-7B/13B} obtain much lower BLEU and CodeBLEU scores in this setting, suggesting that reconstructing a middle span is more demanding than simply continuing code from the left context.
FIM therefore requires not only local code fluency but also tighter consistency with both surrounding contexts.

Another notable pattern is that CodeBLEU is consistently higher than BLEU across models.
Because CodeBLEU additionally incorporates syntax-aware and data-flow matching, many generated solutions appear to recover part of the structural organization of the target code even when token-level overlap is limited.
Overall, the FIM setting provides richer contextual information, but accurate reconstruction of scientific code remains difficult because the missing span encodes domain-specific logic, variable interactions, and project-specific conventions that must fit both surrounding contexts simultaneously.

\vspace{-0.3cm}
\paragraph{Failure Patterns.}
To understand why instruction-tuned models achieve lower BLEU scores on completion tasks, we identify two dominant failure modes. First, instruction-tuned models tend to generate generic boilerplate rather than project-specific continuations, defaulting to statistically common patterns instead of inferring project-internal conventions. Second, they tend to stop early and miss file-level coherence constraints, completing only the immediate logical unit while omitting definitions that are already referenced in the prefix. Detailed examples are provided in Appendix~\ref{app:qualitative}.

% \vspace{-2mm}
% \subsubsection{Answer to RQ3}

% \input{tables/rq4}

\noindent\textbf{\ding{184} Executable Benchmark.}
Table~\ref{tab:rq4} reports the Pass@1 and Pass@5 results on our executable benchmark. 
Overall, the executable benchmark reveals that current models still struggle to generate functionally correct scientific code from natural-language instructions.
Most models achieve only limited success.
Among all evaluated models, \textit{\chatgpt-5.4-mini} performs best, reaching 12.30\% Pass@1 and 15.50\% Pass@5, followed by \textit{o3-mini}.
% with 8.30\% Pass@1 and 10.50\% Pass@5. 
All \textit{StarCoder2} models obtain scores close to 0\% on both metrics. 
We attribute this mainly to these base code completion models are not designed to follow natural-language task instructions directly.
Notably, even \textit{DeepSeek-Coder-6.7B}, which achieves 16.30\% on DSCodeBench, reaches only 6.10\% Pass@1 on our benchmark.
This gap indicates that our benchmark is substantially more challenging than prior data-science benchmarks.

\noindent\textbf{Failure Patterns.} Examining the failure patterns (detailed in Appendix~\ref{sec:app_failure_analysis}), we find notable differences across model families. \textit{StarCoder2} models fail predominantly due to syntax errors, suggesting that non-instruction-tuned code models struggle in our instruction-following setting. By contrast, other models fail more frequently due to execution errors, test failures, incomplete implementations, or missing fields. These indicate that scientific code generation demands syntactic correctness, structural completeness, and precise functional alignment, which no current model reliably achieves.

\begin{table}[h]
\centering
\vspace{-0.5cm}
\caption{\textbf{Left}: Continued pretraining results on F1 evaluated on the PTS completion task. \textbf{Right}: Instruction tuning results on F3/F4 evaluated on the executable benchmark.}
\label{tab:corpus_utility}
\begin{minipage}{0.48\textwidth}
    \centering
    \scriptsize
    \begin{tabular}{lrrr}
    \toprule
    \textbf{Metric} & \textbf{Base} & \textbf{5-epoch} & \textbf{Gain} \\
    \midrule
    BLEU-4    & 0.0228 & 0.1104 & $\times$4.84 \\
    CodeBLEU       & 0.0867 & 0.2463 & $\times$2.84 \\
    Exact Match    & 0.0010 & 0.0200 & $\times$20.00 \\
    \bottomrule
    \end{tabular}
\end{minipage}
\hfill
\begin{minipage}{0.48\textwidth}
    \centering
    \scriptsize
    \begin{tabular}{lrr}
    \toprule
    \textbf{Model} & \textbf{Pass@1} & \textbf{Pass@5} \\
    \midrule
    Qwen2.5-Coder-0.5B (base) & 1.90\% & 5.00\% \\
    Qwen2.5-Coder-0.5B (ours) & 9.10\% & 13.00\% \\
    \bottomrule
    \end{tabular}
\end{minipage}
\vspace{-0.3cm}
\end{table}

\noindent\textbf{\ding{185} Corpus Utility Evaluation}
To demonstrate that \toolname{} provides effective training signal beyond serving as an evaluation resource, we conduct two lightweight training experiments validating the utility of the pretraining corpus (F1) and the instruction-level data (F3/F4) respectively.

\textbf{Pretraining Corpus Utility (F1).}
We perform 5-epoch continued pretraining of GPT-2 (124M) on F1, using a maximum sequence length of 1024 and adjusting the prefix-to-suffix evaluation to a 768-token prefix with a 256-token suffix. As shown in Table~\ref{tab:corpus_utility} (left), continued pretraining yields substantial improvements across all metrics, with CodeBLEU improving by $\times$2.84 and Exact Match by $\times$20, demonstrating that F1 provides effective training signal for scientific code completion.

\textbf{Instruction Tuning Utility (F3/F4).}
We perform instruction tuning of Qwen2.5-Coder-0.5B on our scientific instruction data and evaluate on the executable benchmark. As shown in Table~\ref{tab:corpus_utility} (right), fine-tuning improves Pass@1 from 1.90\% to 9.10\% ($\times$4.79) and Pass@5 from 5.00\% to 13.00\% ($\times$2.60), confirming that F3/F4 provides effective training signal for scientific code generation. Both experiments use small models due to resource constraints; large-scale training experiments remain an important direction for future work.

\vspace{-3mm}
\section{Conclusion and Future Work}
\label{sec:conclusion}
\vspace{-0.2cm}
We present a scientific code corpus constructed from public repositories and organized into four complementary data formats, linking repository-level intent with instruction-level supervision.
Based on this corpus, we evaluate a diverse set of LLMs on three realistic tasks: prefix-to-suffix completion, fill-in-the-middle infilling, and executable code generation.
Our results reveal that scientific code generation remains a challenge for current models.
Across tasks, even the strongest models achieve only limited CodeBLEU on completion and low Pass rates on executable generation.
These findings suggest that future scientific code models must better leverage repository context and external dependencies to generate correct implementations.
Our goal is to support research on reliable AI assistants for scientific software development.

\noindent\textbf{Limitations.} While \toolname{} provides the largest scientific code corpus to date, the corpus may underrepresent certain scientific fields or specialized software ecosystems. Detailed discussion is provided in Appendix~\ref{sec:limitaion}.

\bibliographystyle{unsrtnat}
\bibliography{custom}

\clearpage
%%%%%%%%%%%%%%%%%%%%%%%%%%%%%%%%%%%%%%%%%%%%%%%%%%%%%%%%%%%%

\appendix

\section{Models Used for Data Construction}
\label{sec:appendix_model_roles}

We use a fixed, role-based assignment of models throughout the construction pipeline. The choice is task-driven: rather than switching models arbitrarily across stages, we assign one model family to each type of operation according to its functional requirements. Table~\ref{tab:model_roles} summarizes the assignment.

\begin{table}[h]
\centering
\caption{Role-based model assignment in the \toolname{} construction pipeline.}
\label{tab:model_roles}
\scriptsize
\begin{tabular}{lll}
\toprule
\textbf{Pipeline Stage} & \textbf{Model} & \textbf{Primary Reason} \\
\midrule
Repository classification \& README summarization (F2) & Qwen3-32B & Long-context, local deployment, cost-effective \\
Embedding-based relevance scoring (F3 ranking) & Qwen3-Embedding & Unified representation space across stages \\
Function description generation (F3) & GPT-5 & High-quality NL abstraction from code \\
Problem-solution pair generation (F4) & GPT-5 & High-quality instruction generation \\
Test harness synthesis (benchmark) & GPT-5.3-Codex & Code-specialized structured generation \\
Scientific relevance verification (F3) & NatureLM & Domain-specialized scientific verification \\
\bottomrule
\end{tabular}
\end{table}

\paragraph{Repository-level classification and summarization.}
We use Qwen3-32B for repository-level relevance classification and \texttt{README} summarization. These stages operate over tens of thousands of repositories and require long-context understanding, stable structured outputs, and practical scalability. Importantly, this stage involves large-scale batch processing where API-based models would be prohibitively expensive; Qwen3-32B is deployed locally, significantly reducing computational cost. According to its technical report~\cite{Yang2025}, Qwen3-32B matches or outperforms OpenAI o3-mini on general-purpose benchmarks, making it a strong and cost-effective choice. Using the same model for both filtering and summarization also reduces representational inconsistency between the two stages.

\paragraph{Embedding-based relevance scoring.}
We use Qwen3-Embedding for all similarity computations in the pipeline, including both repository-level and function-level relevance scoring. The key reason is consistency: using a single embedding model ensures that all similarity scores are computed in the same representation space and remain directly comparable across stages.

\paragraph{Instruction and code generation.}
For higher-precision generative stages, we use the GPT-5 family. GPT-5 is used for function-level description generation (F3) and problem-solution pair generation (F4), as these stages require high-quality natural language abstraction from code, dependencies, and repository context. For executable benchmark construction, we use GPT-5.3-Codex for test harness synthesis, since this stage directly requires structured Python code generation of setup code and test assertions. Using the same model family across generation stages improves consistency, while reserving the code-specialized variant for executable test construction.

\paragraph{Scientific relevance verification.}
We use NatureLM~\cite{Xia2025} as an automatic verifier for F3 function descriptions. NatureLM is a domain-specialized scientific language model, making it particularly well-suited for assessing whether generated descriptions are both consistent with the provided context and relevant to the target scientific domains — a judgment that benefits from domain-specific pretraining beyond general-purpose LLMs.

\textbf{Instruction and code generation.}
For higher-precision generative stages, we use the GPT-5 family.
GPT-5 is used for function-level and problem-level instruction generation because these stages require high-quality natural-language abstraction from code, dependencies, and repository context.
For executable benchmark construction, we use GPT-5.3-Codex for test synthesis, since this stage directly requires structured code generation of Python setup code and test harnesses.
Using the same model family across generation stages improves consistency, while reserving the code-specialized variant for executable test construction.

Overall, this model assignment is based on the functional requirements of each stage: long-context repository processing, unified embedding-based scoring, and high-precision text/code generation.
We do not claim that this configuration is uniquely optimal; rather, it provides a consistent and practical setup for constructing the corpus and benchmark at scale.

\section{Stage-wise Quality Control}
\label{sec:appendix_quality}

To ensure the reliability and faithfulness of the constructed corpus and benchmark, we apply explicit quality control mechanisms at each generative stage of the pipeline. Table~\ref{tab:quality_control} provides an overview of the measures applied at each stage.

\begin{table}[h]
\centering
\caption{Stage-wise quality control mechanisms in the \toolname{} construction pipeline.}
\label{tab:quality_control}
\small
\begin{tabular}{lll}
\toprule
\textbf{Stage} & \textbf{Model} & \textbf{Quality Control Mechanism} \\
\midrule
Repository retrieval (F1) & Qwen3-32B & Binary relevance judgment with reasoning; \\
& & repositories with $<$10 stars excluded \\
README summarization (F2) & Qwen3-32B & Fixed JSON schema constrains output format \\
Function selection (F3) & Qwen3-Embedding & Dual-level Finalscore filtering \\
& & ($\text{mdscore} \times \text{codescore}$) \\
Function description (F3) & GPT-5 + NatureLM & NatureLM verifies consistency and \\
& & scientific relevance of generated descriptions \\
Problem-solution pairs (F4) & GPT-5 & Real code snippets as anchors; \\
& & solutions constrained to provided dependencies \\
Test harness synthesis & GPT-5.3-Codex & End-to-end runnable verification under \\
(benchmark) & & sandboxed isolation with timeout control \\
Benchmark validation & Human & Manual review of 40 sampled tasks (20\%); \\
& & 85\% full alignment, Cohen's $\kappa \approx 0.92$ \\
\bottomrule
\end{tabular}
\end{table}

\paragraph{Repository retrieval.}
Repository quality is enforced through two complementary filters. First, we exclude repositories with fewer than 10 stars, following prior work~\cite{Sun2023, Wen2019}, to remove low-quality or inactive projects. Second, Qwen3-32B performs a binary relevance judgment (YES/NO) with explicit reasoning for each repository, assessing whether the \texttt{README} describes a project that actually implements computational science methods rather than merely mentioning related keywords. Together, these filters reduce the initial candidate pool of 1,311,568 repositories to 37,737 high-quality, domain-relevant projects.

\paragraph{README summarization.}
F2 summaries are generated under a fixed JSON schema specifying five structured fields: project overview, main features, dependencies, scientific computing characteristics, and typical input/output information. The schema constraint prevents free-form generation and ensures that all summaries are structured, comparable, and grounded in the original \texttt{README} content.

\paragraph{Function selection and relevance scoring.}
Functions in F3 are selected based on $Final_{score} = md_{score} \times code_{score}$), where $md_{score}$ measures the embedding similarity between the domain keyword set and the repository \texttt{README}, and $code_{score}$ measures the similarity between the same keyword set and the function body. This dual-level scoring ensures that selected functions are scientifically grounded at both the repository intent level and the implementation level, filtering out functions that are incidentally present in scientific repositories but do not themselves encode scientific logic.

\paragraph{Function description generation.}
Generated F3 descriptions are subject to automatic verification by NatureLM~\cite{Xia2025}, a domain-specialized scientific language model. NatureLM assesses each generated description along two dimensions: (1) consistency with the provided code and repository context, and (2) relevance to the target scientific domains. Descriptions that fail either check are treated as low-confidence and filtered out before entering the dataset. To directly quantify the noise level of the remaining descriptions, we conducted a human evaluation on a random sample of 50 F3 instances, assessed by two independent annotators:

\begin{table}[h]
\centering
\caption{Human evaluation results on 50 randomly sampled F3 function descriptions.}
\label{tab:f3_human_eval}
\small
\begin{tabular}{lr}
\toprule
\textbf{Metric} & \textbf{Result} \\
\midrule
Faithfulness rate & 92\% \\
Hallucination rate & 6\% \\
Inter-annotator agreement (Cohen's $\kappa$) & 0.82 \\
\bottomrule
\end{tabular}
\end{table}

The 92\% faithfulness rate and low hallucination rate of 6\% confirm that the NatureLM verification step effectively filters low-quality generations, and that the retained F3 descriptions reliably reflect the underlying code.

\paragraph{Problem-solution pair generation.}
F4 construction follows the OSS-Instruct principle~\cite{Wei2023}, using real code snippets extracted from F3 functions as anchors. Generated solutions are explicitly constrained to dependencies present in the provided repository context, reducing the risk of hallucinated library usage. This grounding in real code ensures that F4 instances reflect genuine scientific computing patterns rather than synthetic constructs disconnected from real workflows.

\paragraph{Executable benchmark construction and validation.}
Test harnesses are synthesized by GPT-5.3-Codex and subsequently executed end-to-end under subprocess-based sandboxed isolation with timeout control. Only instances where the reference implementation passes all synthesized tests without runtime errors are retained, ensuring that every benchmark task provides a reliable and reproducible pass/fail signal. A detailed manual audit of the retained benchmark instances is reported in Appendix~\ref{sec:app_test_audit}.

\section{Verification of Executable Benchmark Test Cases}
\label{sec:app_test_audit}

To ensure that the test harnesses in our executable benchmark faithfully verify functional correctness, we conduct a manual analysis of a subset of tasks. 
The aim is to assess whether the automatically generated test cases 
1) accurately reflect the specification described in the prompt and the reference implementation and 
2) can be executed in a sandboxed environment without dependency issues.
Our manual analysis is inspired by best practices in prior work on code generation benchmarks, such as MBPP~\cite{Austin2021} and HumanEval~\cite{Chen2021}.

\textbf{Sampling and Procedure.}
Out of the 200 benchmark tasks, we randomly select 40 tasks (20\%) for manual review. 
This sampling ratio balances coverage with feasibility.
Two independent reviewers with domain expertise in computational science and software engineering carry out the review. 
Each task is reviewed by both reviewers to measure agreement.
For each selected task, the reviewers examine:
1) \textit{Prompt–test alignment}: Whether the test suite covers all major functional requirements described in the natural-language prompt.
2) \textit{Reference compatibility}: Whether the reference implementation provided passes the test suite without modification.
3) \textit{Sandbox executability}: Whether the test suite can run successfully in the provided environment, including proper setup of stubs or mocks.
4) \textit{Metrics recorded}: We record the number of tasks where (i) tests fully align with the prompt, (ii) minor issues exist (\eg missing edge-case assertions), or (iii) significant problems are found (\eg test errors or misinterpretations).

\textbf{Results of the Manual Review.}
Among the 40 reviewed tasks, 34 (85\%) had test suites that perfectly matched the prompt and accepted the reference implementation without issue. 
6 tasks (15\%) exhibited minor issues, such as missing corner-case checks or overly permissive assertions; these could be addressed by adding or tightening a few assertions.
All 40 tasks could be executed end-to-end within the benchmark’s sandbox environment.
A few required minor adjustments to stub functions (\eg adding a missing import), which are subsequently incorporated into the benchmark.
The two reviewers agreed on the classification of 38 out of 40 tasks (Cohen's $\kappa \approx 0.92$)~\cite{Cohen1960}, indicating a high level of consistency in the audit process.

\section{Keyword List for Repository Retrieval}
\label{sec:keyword_list}

Table~\ref{tab:keywords} lists the 198 domain keywords used to retrieve scientific repositories from GitHub.

\begin{table*}[t]
\centering
\tiny
\caption{Scientific domain keywords used for repository retrieval (198 terms).}
\label{tab:keywords}
\setlength{\tabcolsep}{1.5pt}
\begin{tabular}{lllll}
\toprule
Chemistry & Biology & Biochemistry & Omics & Medicine \\
Pharmacology & Toxicology & Bioinformatics & Bioengineering & Biophysics \\
Viral & Microbial & Protein & Gene & DNA \\
RNA & Vaccine & Computational Biology & Computational Biochemistry & Computational Chemistry \\
Computational Materials & Quantum Chemistry & Disease & Biomedical & Material \\
Pharmacogenetics & Pharmacogenomics & Modeling & Networks & In Silico \\
Pathology & Physiology & Genomics & Proteomics & Transcriptomics \\
Metabolomics & Glycomics & Lipidomics & Immunology & Microbiology \\
Molecular biology & Pharmaceutics & Network pharmacology & Epigenetics & Sequencing \\
Multi-omics & Biomarker & System biology & Synthetic biology & Cell biology \\
Cancer biology & Ensemble & Personalized & Lipid & Metabolic \\
Genesis & Ion & Heterogeneity & Generative & Receptor \\
Ligand & Organoid & Evolution & Pathogens & Homeostasis \\
Allele & Genotype & Phenotype & Antibody & Antigen \\
Nucleic acids & Carbohydrate & Substrate & Inhibition & Activation \\
Allosteric & Cofactor & Coenzyme & Enzyme & Redox \\
Hydrophilic & Hydrophobic & Codon & Transcription & Translation \\
Pathway & Cycle & Signaling & Dynamics & Kinetics \\
Docking & Spectrometry & Profiling & Diagnostics & CRISPR \\
Pharmacokinetics & Pharmacodynamics & Absorption & Mechanism of action & Agonist \\
Antagonist & Bioavailability & Half-life & Reaction & Drug \\
Biologics & Pharmacometrics & Beta-blocker & Regulatory networks & Multi-scale modeling \\
Single-cell & Spatial biology & Monte Carlo & System immunology & Metagenomics \\
QSAR & QAPR & Chemical space & AlphaFold & Folding \\
Digital twin & Virtual human & Gene editing & Bio foundation model & Biotechnology \\
Assay & Lead discovery & High-throughput & Screening & Hit-to-lead \\
Lead optimization & De novo & ADMET & Translational medicine & Drug repurpose \\
Conjugate & Agent-based model & Compartmental model & Reproduction number & Nowcasting \\
Phylodynamic model & Physiologically based pharmacokinetics model & PBPK model & Organ-on-a-chip & Anomaly detection \\
Stochastic modeling & Genomic surveillance & Antimicrobial resistance modeling & AMR & Pandemic \\
Digital PCR & Next-generation sequencing & Biosensors & Imaging & Sensors \\
Quantum mechanics & DFT & Ab initio & Hartree-Fock & Coupled cluster \\
Electronic structure & Homo-Lumo & Conformation & Cheminformatics & QM/MM \\
First-principles based DFT & Diffusion & Finite element method & Phase-field technique & Potential \\
Metamaterial & 2D & 3D & Porous & Crystal \\
Rosettafold & Gene regulatory networks & Cell atlas & Human atlas & Spatial transcriptomics \\
Pseudotime analysis & Quantum biology & Metabolic flux analysis & Free energy perturbation & Protein-protein \\
Explainable AI & Neurology & Reinforcement learning & Generative AI & Flow matching \\
Generative adversarial networks & GAN & Variational autoencoders & VAE & Autoregressive \\
Transformer & Recurrent neural networks & RNN &  &  \\
\bottomrule
\end{tabular}
\end{table*}

\section{Qualitative Examples of Completion Task Failures}
\label{app:qualitative}

We present two representative examples illustrating why instruction-tuned models achieve lower BLEU scores than code-specialized base models on completion tasks.

\subsection*{Example 1: Fallback to Generic Boilerplate}

This example is drawn from the \texttt{chemprop} chemistry project. The prefix ends with an incomplete \texttt{app.config.from\_object} call, requiring the model to infer the project-specific configuration module and subsequent initialization logic.

\begin{lstlisting}[language=Python, caption={Prefix}, label={lst:ex1_prefix}]
"""Runs the web interface version of chemprop, allowing for 
training and predicting in a web browser."""
import os
from flask import Flask

app = Flask(__name__)
app.config.from_object
\end{lstlisting}

\begin{lstlisting}[language=Python, caption={Ground Truth}, label={lst:ex1_gt}]
('chemprop.web.config')

os.makedirs(app.config['CHECKPOINT_FOLDER'], exist_ok=True)
os.makedirs(app.config['DATA_FOLDER'], exist_ok=True)

from chemprop.web.app import views
\end{lstlisting}

\begin{lstlisting}[language=Python, caption={Instruction-tuned Model Generation}, label={lst:ex1_gen}]
('config.DefaultConfig')

if __name__ == '__main__':
    app.run(host='0.0.0.0', port=5000)
\end{lstlisting}

The ground truth passes the project-specific module \texttt{chemprop.web.config}, creates project-internal directories, and imports project routes. The instruction-tuned model instead generates \texttt{config.DefaultConfig} followed by a standard \texttt{app.run(...)} pattern common in Flask tutorials but entirely disconnected from the codebase. This illustrates that instruction-tuned models default to statistically common continuations rather than project-specific ones, as they lack access to the broader codebase context needed to infer internal module names and directory conventions.

\subsection*{Example 2: Early Stopping and Missing File-level Coherence}

This example is drawn from a Django REST framework project. The prefix already references three custom exception classes (\texttt{InvalidFilters}, \texttt{InvalidData}, \texttt{BadRequest}) and truncates mid-function, requiring the model to complete the current branch and define the referenced classes.

\begin{lstlisting}[language=Python, caption={Prefix (truncated)}, label={lst:ex2_prefix}]
from rest_framework.exceptions import APIException
from rest_framework.views import exception_handler

def custom_exception_handler(exc, context):
    response = exception_handler(exc, context)
    if response is not None:
        data = {"error_type": "", "message": "", "details": ""}
        if isinstance(exc, (InvalidFilters, InvalidData, BadRequest)):
            data["error_type"] = exc.default_code
            data["message"] = exc.detail
            data["details"] = exc.details
        if isinstance(exc, ValidationError):
            detail = exc.detail
            if isinstance(detail, dict):
                errors = []
                for key, value in detail.items():
                    errors.append({...})
                if len(errors) > 1:
                    ... # truncated here
\end{lstlisting}

\begin{lstlisting}[language=Python, caption={Ground Truth}, label={lst:ex2_gt}]
                else:
                    data = errors[0]
                response.data = data
    return response

class InvalidFilters(APIException):
    status_code = 400
    default_code = "invalid_filters"
    def __init__(self, message=None, invalid_filters=None):
        super().__init__(detail=message)
        self.details = invalid_filters

class InvalidData(APIException):
    status_code = 400
    default_code = "invalid_data"
    def __init__(self, message=None, details=None):
        super().__init__(detail=message)
        self.details = details

class BadRequest(APIException):
    status_code = 400
    default_code = "bad_request"
    def __init__(self, message=None, error_type=None):
        super().__init__(detail=message)
\end{lstlisting}

\begin{lstlisting}[language=Python, caption={Instruction-tuned Model Generation}, label={lst:ex2_gen}]
                else:
                    data.update(errors[0])  # semantic error: should be data = errors[0]
                response.data = data
    return response
    # stops here -- three exception classes never generated
\end{lstlisting}

The model completes only the immediate else branch with a semantic error (\texttt{data.update(errors[0])} instead of \texttt{data = errors[0]}) and stops at \texttt{return response}, generating none of the three exception class definitions. Yet \texttt{InvalidFilters}, \texttt{InvalidData}, and \texttt{BadRequest} are already referenced in the prefix, a strong signal that they must be defined in the same file. This reveals that instruction-tuned models treat completion as a local task and stop once the immediate logical unit is finished, missing the file-level coherence constraint that base code models handle more naturally through next-token prediction.

\section{Failure Analysis on the Executable Benchmark.}
\label{sec:app_failure_analysis}

\begin{figure*}[t]
    \centering
    \includegraphics[width=0.8\linewidth]{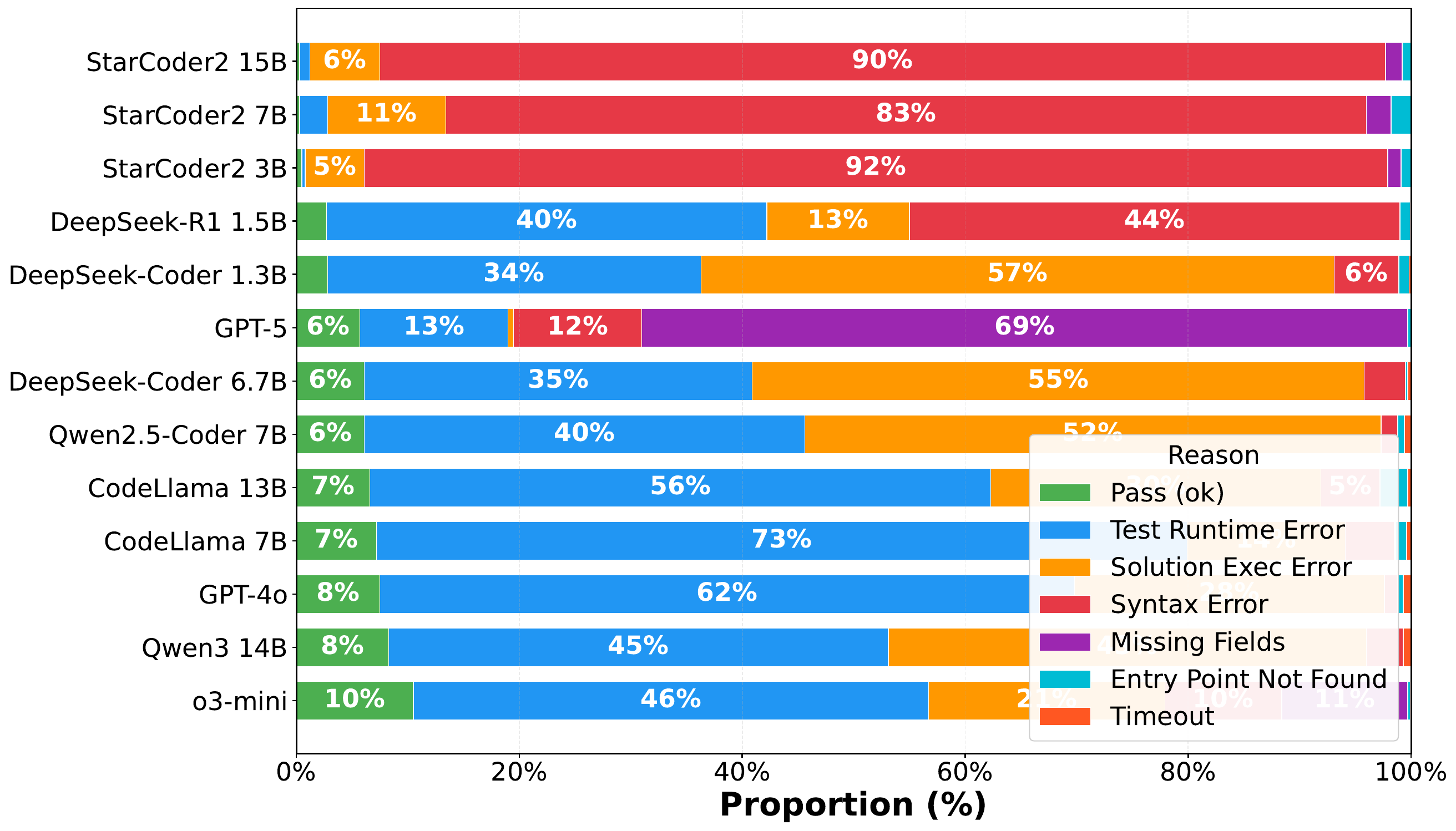}
    \caption{Failure breakdown of different models on the executable benchmark. 
    Each bar shows the proportion of outcomes categorized by failure reason, including syntax errors, runtime errors during test execution, missing fields in the generated implementation, incorrect entry points, setup execution errors, and timeouts.}
    \label{fig:failure_analysis}
\end{figure*}

Figure~\ref{fig:failure_analysis} presents a breakdown of failure types on the executable benchmark.
A clear pattern is that the dominant failure reason differs substantially across model families.
For the \textit{StarCoder2} models, failures are overwhelmingly dominated by \textit{syntax errors}, accounting for about 85\%--87\% of all cases, which is consistent with their lack of instruction tuning in this instruction-following setting.
By contrast, many other open-source code models fail mainly due to \textit{solution execution errors} or \textit{test runtime errors}.
For example, \textit{DeepSeek-Coder-1.3B}, \textit{DeepSeek-Coder-6.7B}, and \textit{Qwen2.5-Coder-7B} show large proportions of solution execution errors, while \textit{CodeLlama-7B}, \textit{CodeLlama-13B}, and \textit{GPT-4o} are dominated by test runtime errors, suggesting that these models more often generate syntactically valid code but still fail to satisfy task requirements during execution.
Closed-source models also exhibit distinct failure modes.
\textit{GPT-5} stands out for having a large proportion of \textit{missing fields} errors, whereas \textit{o3-mini} and \textit{Qwen3-14B} mainly fail through runtime errors after successful generation.
Overall, the results suggest that executable scientific code generation requires not only syntactic correctness, but also structurally complete implementations and precise functional alignment with the benchmark specification.

\section{Additional Information of the Code Corpus}

% \begin{wrapfigure}{r}{0.49\textwidth}
%   \centering
%   \begin{subfigure}[t]{0.49\textwidth}
%     \centering
%     \includegraphics[width=\linewidth]{image_10.png} 
%         \caption{Keyword word cloud for repository search queries (Stage~1).}
%         \label{fig:domain}  
%   \end{subfigure}
%   \begin{subfigure}[t]{0.49\textwidth}
%     \centering
%     \includegraphics[width=\linewidth]{image_9.png} 
%     \caption{Top-20 file extension distribution across all collected repositories.}
%     \label{fig:distribution}
%   \end{subfigure} 
%   \caption{Overview of the data collection. (a) Keyword word cloud of repository search queries in Stage~1. (b) Top-20 file extension distribution across the collected repositories.}
% \end{wrapfigure}

\subsection{Scientific Domain Coverage}

\begin{figure*}[t]
    \centering
    \includegraphics[width=0.8\linewidth]{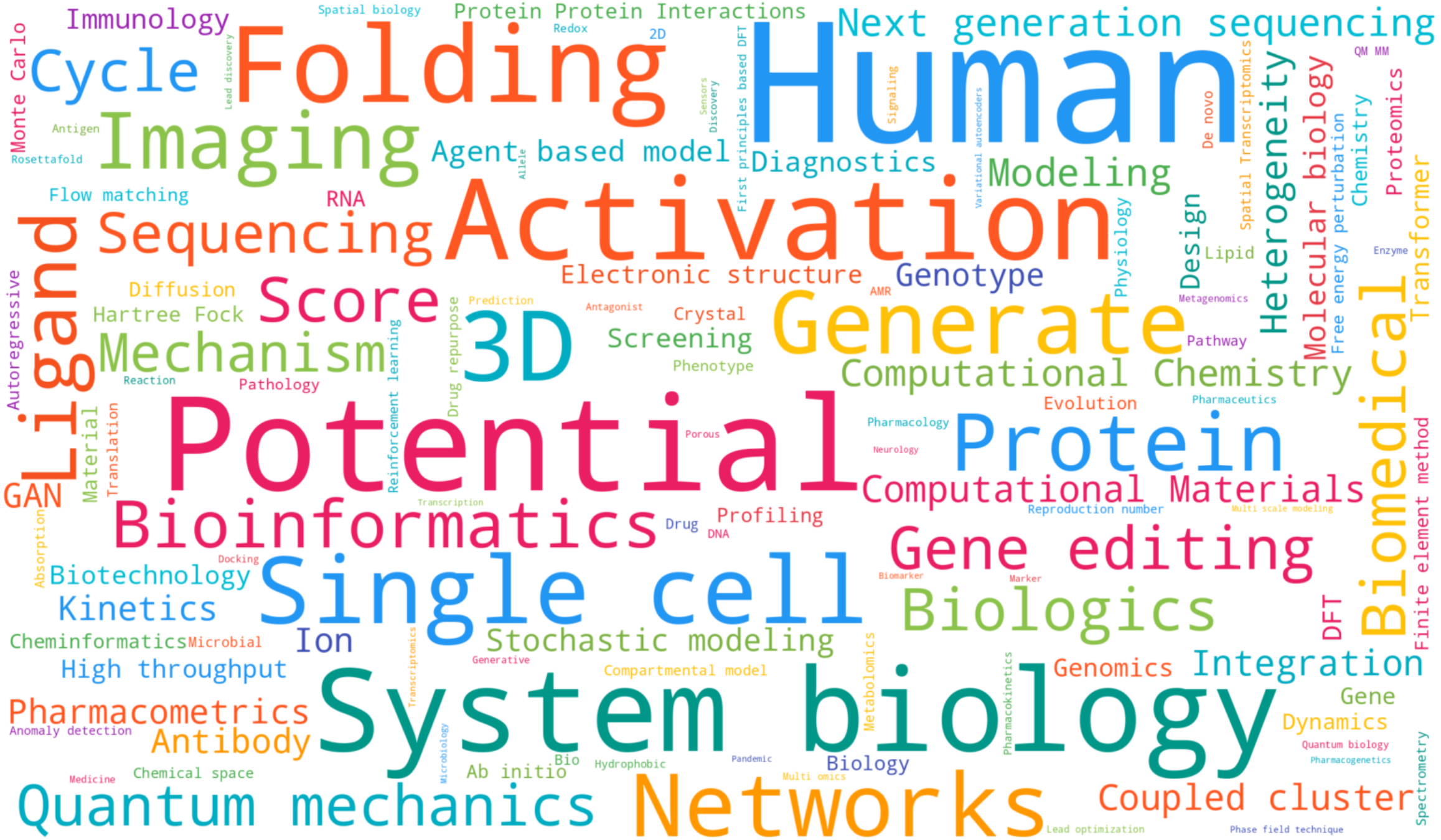}
    \caption{Word cloud of scientific keywords extracted from the collected repositories. 
The size of each term reflects its frequency in the dataset, illustrating the diverse scientific domains represented in the corpus.}
    \label{fig:keyword_wordcloud}
\end{figure*}

To examine the topical diversity of the collected repositories, we analyze the distribution of domain-related keywords extracted from the dataset.
Figure~\ref{fig:keyword_wordcloud} visualizes these keywords using a word cloud, where the size of each term is proportional to its frequency in the corpus.
Several major scientific themes can be observed.
Keywords such as \textit{Human}, \textit{Potential}, \textit{Folding}, \textit{Activation}, and \textit{System biology} appear prominently, indicating that the collected repositories cover a wide range of computational tasks in life sciences and molecular modeling.
At the same time, terms such as \textit{Protein}, \textit{Sequencing}, \textit{Gene editing}, and \textit{Bioinformatics} highlight the strong presence of modern data-driven biological research.
In addition to biology-related topics, the corpus also includes computational themes such as \textit{3D}, \textit{Networks}, and \textit{Quantum mechanics}, suggesting that the dataset spans multiple scientific disciplines and computational paradigms.
Overall, the keyword distribution demonstrates that the collected corpus covers a broad spectrum of scientific research areas, providing a diverse foundation for constructing the executable benchmark.

\begin{figure}[t]
  \centering
  \begin{subfigure}[b]{0.48\linewidth}
    \centering
    \includegraphics[width=\linewidth]{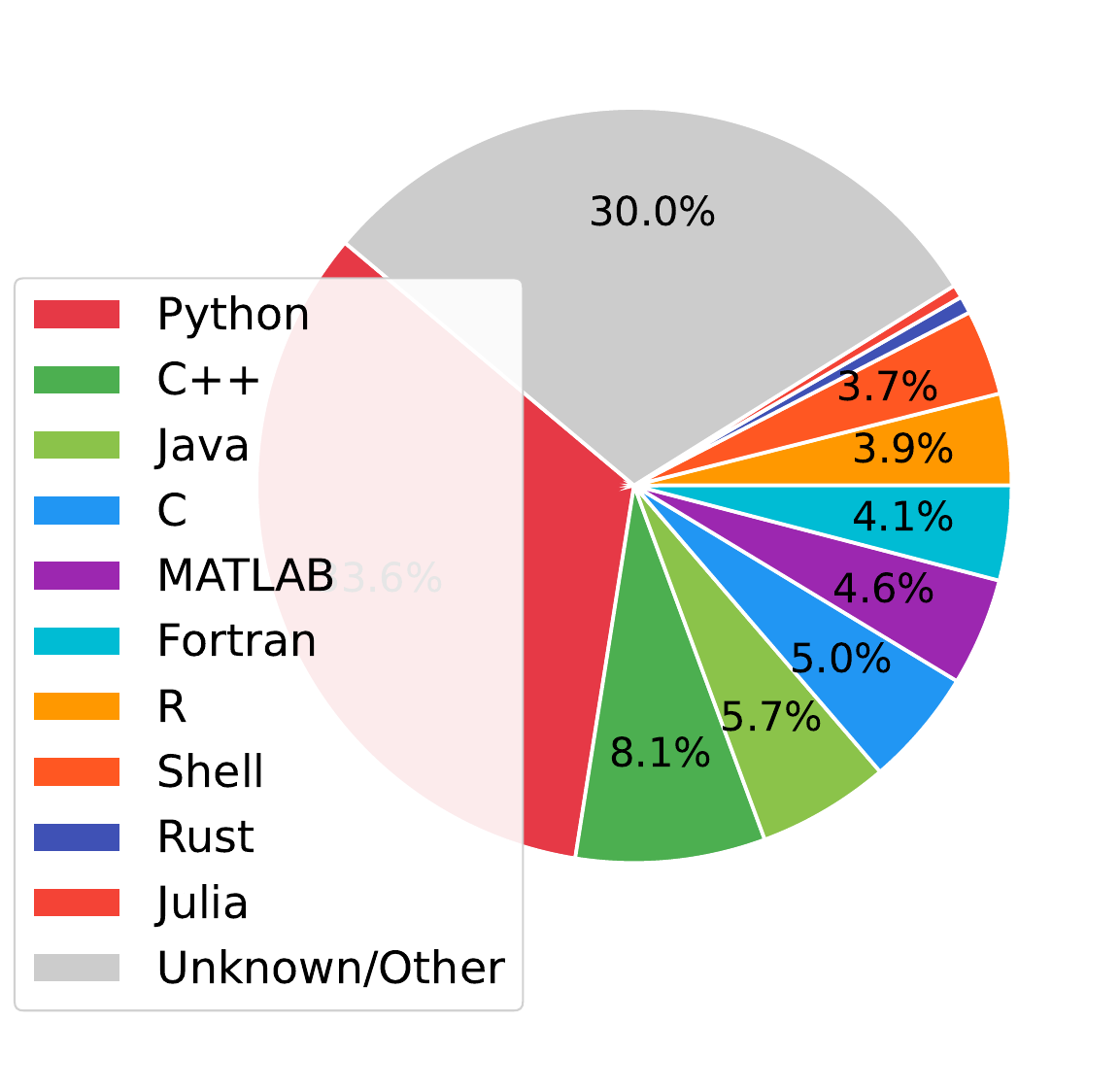}
    \caption{The proportion of files by language.}
  \end{subfigure}
  \hfill
  \begin{subfigure}[b]{0.48\linewidth}
    \centering
    \includegraphics[width=\linewidth]{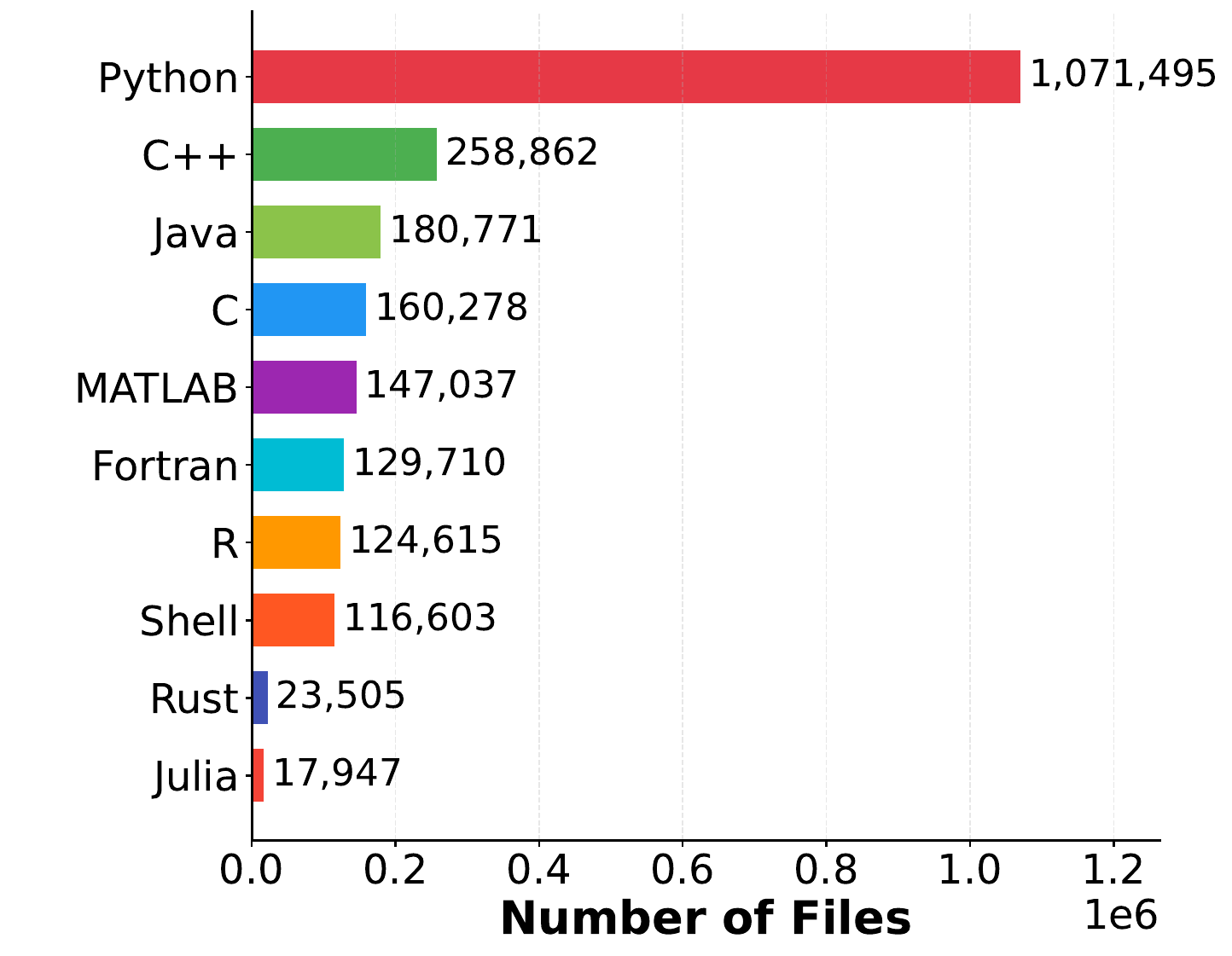}
    \caption{The absolute number of files.}
  \end{subfigure}
    \caption{Programming-language distribution in the collected scientific code corpus.}
  \label{fig:language_distribution}
  \vspace{-0.6cm}
\end{figure}

\subsection{Programming-Language Distribution}

We analyze the programming-language composition of the collected scientific code corpus.
Figure~\ref{fig:language_distribution} presents both the percentage share and the absolute number of files for each language.

Python dominates the dataset, accounting for the largest proportion of files by a substantial margin.
This observation aligns with the widespread adoption of Python in modern scientific computing, particularly in data analysis, machine learning, and bioinformatics workflows.
At the same time, several traditional scientific computing languages remain prominent.
Languages such as C, C++, and Fortran contribute a significant number of files, reflecting their continued use in performance-critical numerical simulations and high-performance computing applications.
MATLAB and R also appear frequently, which is consistent with their popularity in statistical modeling and scientific data analysis.

Overall, the corpus exhibits a diverse programming-language composition that spans both modern scripting languages and traditional high-performance computing languages, providing a realistic environment for evaluating and training scientific large language models.

\begin{figure*}[t]
    \centering
    \includegraphics[width=\linewidth]{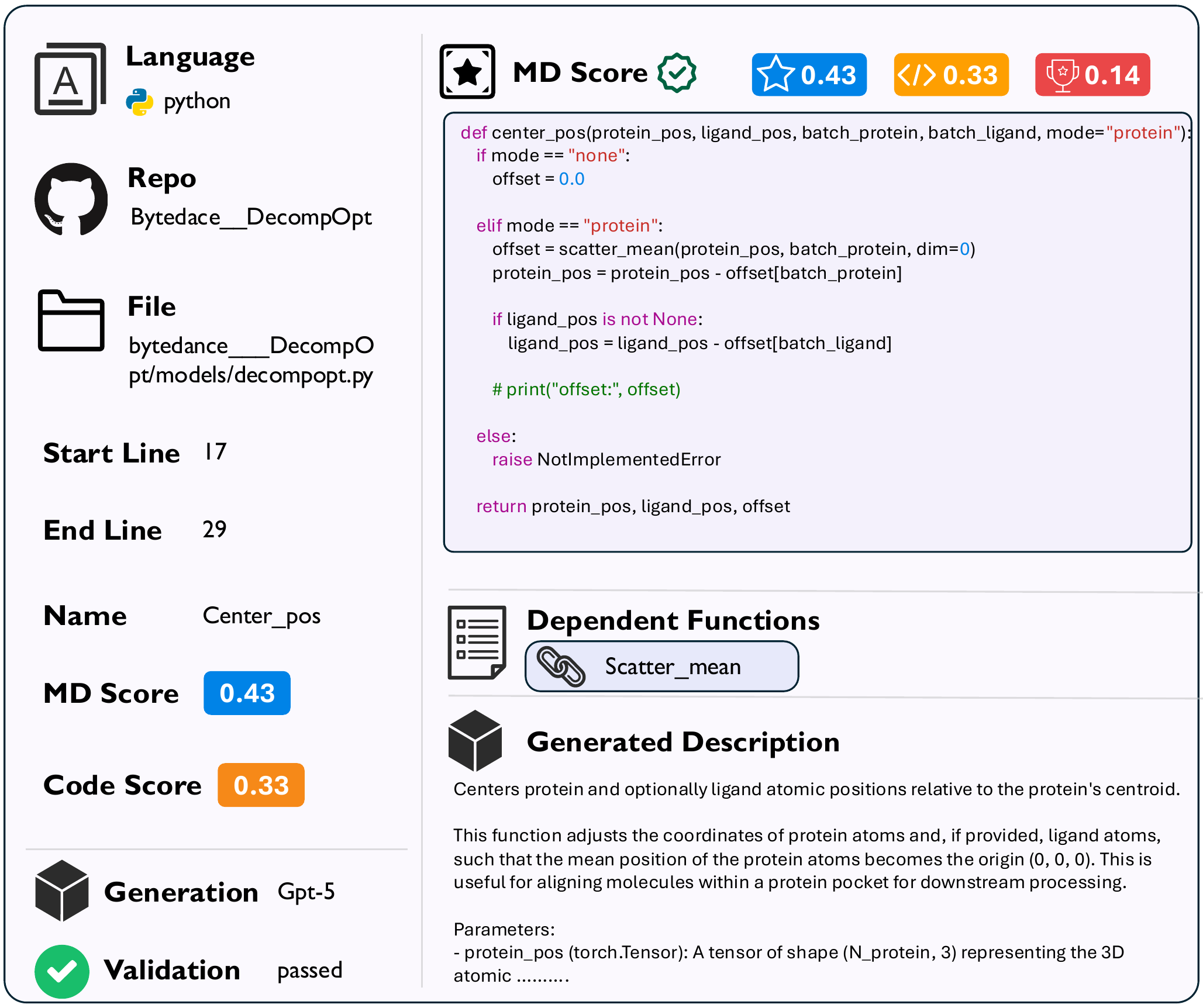}
    \caption{An simplified example in Function Instruction dataset.}
    \label{fig:instruction}
\end{figure*}

\begin{figure*}[t]
    \centering
    \includegraphics[width=\linewidth]{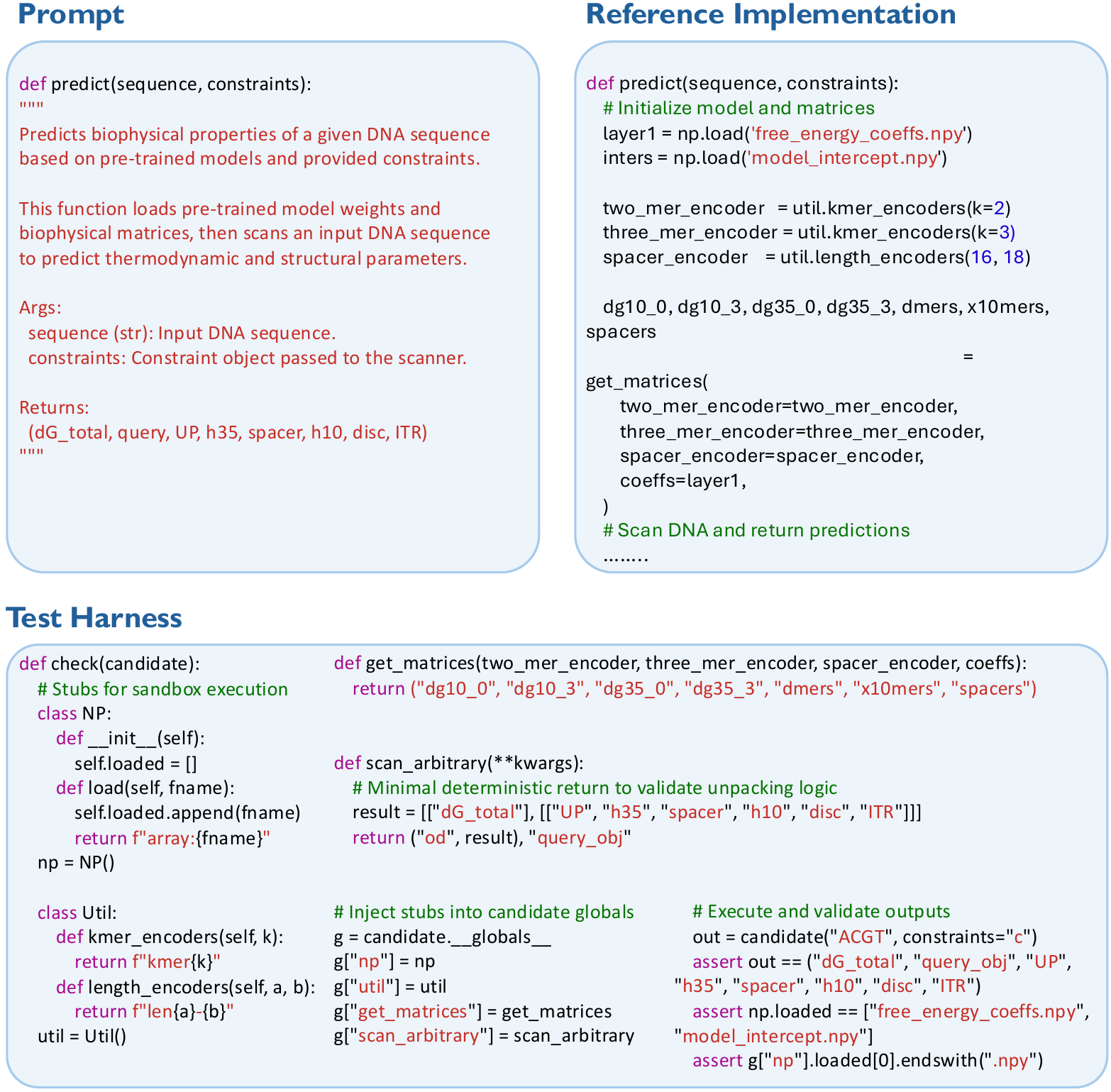}
    \caption{An simplified example in executable benchmark.}
    \label{fig:benchmark_example}
\end{figure*}

\section{F3: Function Instruction Example}
\label{sec:format3-example}

Figure~\ref{fig:instruction} shows a representative F3 (Function Instruction) instance.
In F3, each sample pairs a real function with an evidence-constrained natural-language description, so that the resulting data can support function retrieval, and instruction tuning under domain constraints. 
Concretely, the example anchors on a Python function \texttt{center\_pos}, and retains full traceability metadata (language, repository, file path, line span, and function name/qualified name), together with the raw function body extracted from F1 via \texttt{tree-sitter} parsing. 
The instance is selected through dual evidence: an $md_{score}$ measuring the embedding similarity between the domain keyword set and the repository \texttt{README}, and a $code_{score}$ measuring the similarity between the same keyword set and the function body (both computed using Qwen3-embedding). 
These signals are combined as $Final_{score}=md_{score}\times code_{score}$, which is also displayed in the example for ranking and filtering. 
For the selected function, we additionally attach lightweight context, including extracted dependencies (\eg the salient call \texttt{scatter\_mean}) and the \texttt{README} summary from F2. 
Finally, we prompt \chatgpt-5 to generate a docstring-style description that must stay within this evidence, and use NATURELM~\cite{Xia2025} to filter low-confidence generations. 
As illustrated, the resulting F3 record exposes not only the function and its description, but also the associated evidence, scores, and validation status, enabling transparent reuse in downstream tasks.

\section{Executable Benchmark Instance}
\label{sec:execute_example}

Figure~\ref{fig:benchmark_example} illustrates a simplified example from our executable benchmark to demonstrate how each task instance is structured and evaluated.

Each benchmark instance consists of three core components: a \textit{Prompt}, a \textit{Reference Implementation}, and a \textit{Test Harness}. 
The prompt provides a natural-language specification of the task together with the target function signature. 
In the example shown in Figure~\ref{fig:benchmark_example}, the prompt describes a function \texttt{predict} that estimates biophysical properties of a DNA sequence using pre-trained models and thermodynamic matrices. 
The description outlines the expected inputs (a DNA sequence and constraint parameters) and the required outputs, which include several predicted thermodynamic quantities such as \textit{dG\_total}, \textit{UP}, and \textit{h35}.

The reference implementation provides a correct solution that satisfies the prompt specification. 
In this example, the implementation first loads pre-trained parameters from external files, initializes encoding utilities, and constructs the necessary matrices used for prediction. 
The sequence is then processed through a scanning procedure that produces multiple thermodynamic metrics, which are finally returned as the output tuple.

To enable automated and reproducible evaluation, each instance is accompanied by a test harness that verifies the functional behavior of generated solutions. 
The test harness executes the candidate implementation inside a sandboxed environment and replaces external dependencies with lightweight stubs. 
For example, the benchmark substitutes simplified implementations of modules such as \texttt{numpy}, \texttt{util}, and helper functions like \texttt{get\_matrices} and \texttt{scan\_arbitrary}. 
These stubs provide deterministic outputs that allow the test harness to verify whether the candidate implementation correctly invokes the expected functions, loads the required model parameters, and returns outputs in the correct format.

After the environment is configured, the harness calls the candidate function with a test DNA sequence and a constraint object. 
Assertions are then applied to ensure that the returned tuple matches the expected prediction structure and that the required model files have been loaded. 
Through this executable evaluation mechanism, benchmark instances assess not only the syntactic validity of generated code but also its functional correctness under controlled execution.

\section{Limitations}
\label{sec:limitaion}
Although \toolname{} provides a large-scale corpus and executable benchmark for scientific code generation, several limitations remain. First, while our repository retrieval pipeline aims to cover a broad range of computational science domains, the collected corpus may still underrepresent certain scientific fields or specialized software ecosystems. Expanding domain coverage by incorporating additional repositories is an important direction for future work. Second, a portion of the instruction-level data (\textbf{F4}) is synthesized from real code using LLM-based generation. Although we apply validation and filtering procedures to ensure consistency with the repository implementations, automatically constructed instructions may still introduce noise or simplifications compared to tasks crafted by experts. Third, the executable benchmark currently contains a limited number of tasks due to the stringent requirements of runnable environments and test harness construction. Scaling executable evaluation while maintaining reliability remains a key challenge for future dataset and benchmark development.

\section{Prompt Templates}
\label{sec:prompt_temp}

\subsection{Prompt Template for F1}
\label{sec:append_f1}

\begin{figure*}[t]
    \centering
    \includegraphics[width=\linewidth]{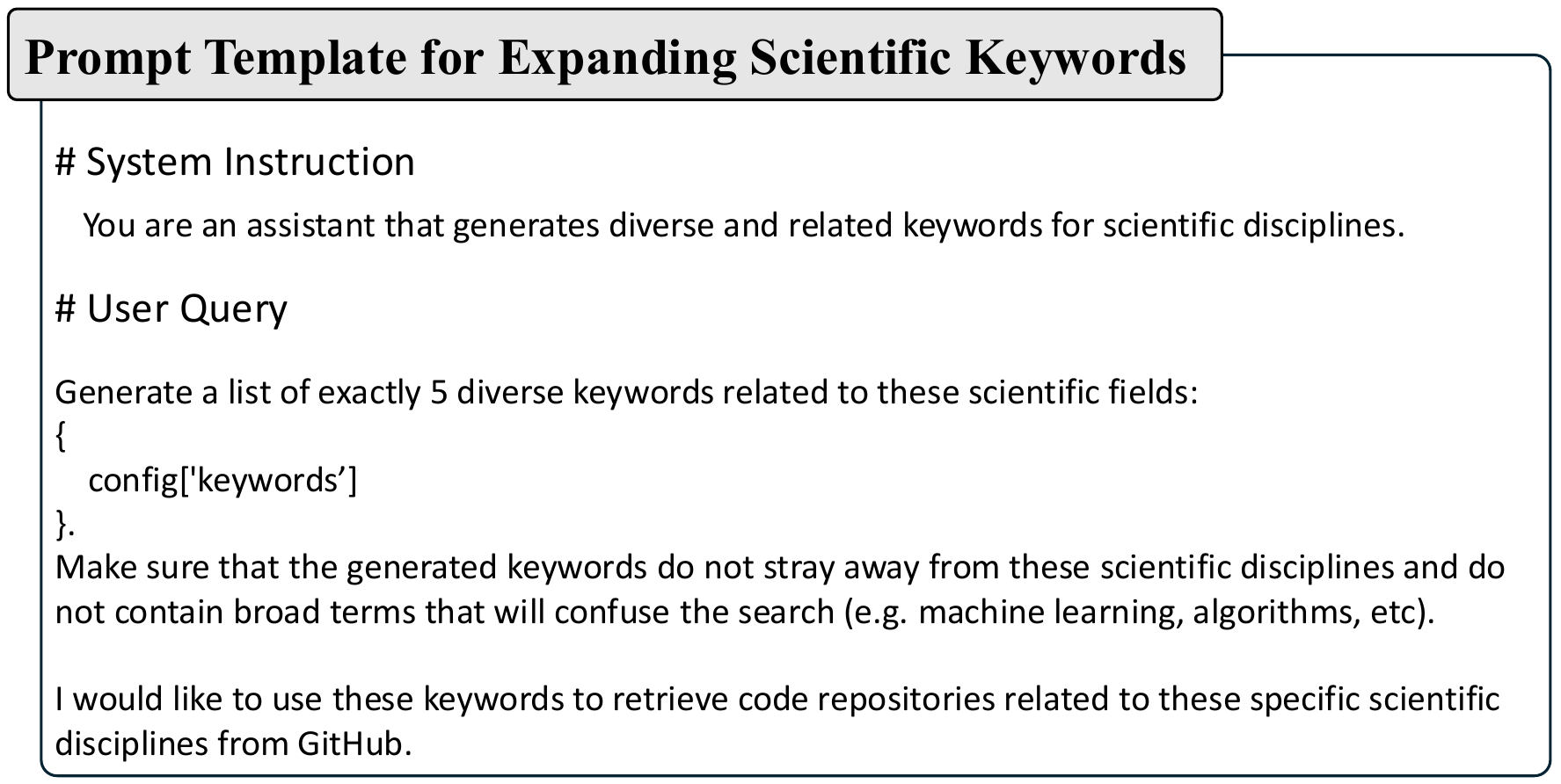}
    \caption{Prompt Template for Expanding Scientific Keywords.}
    \label{fig:prompt1}
\end{figure*}

\begin{figure*}[t]
    \centering
    \includegraphics[width=\linewidth]{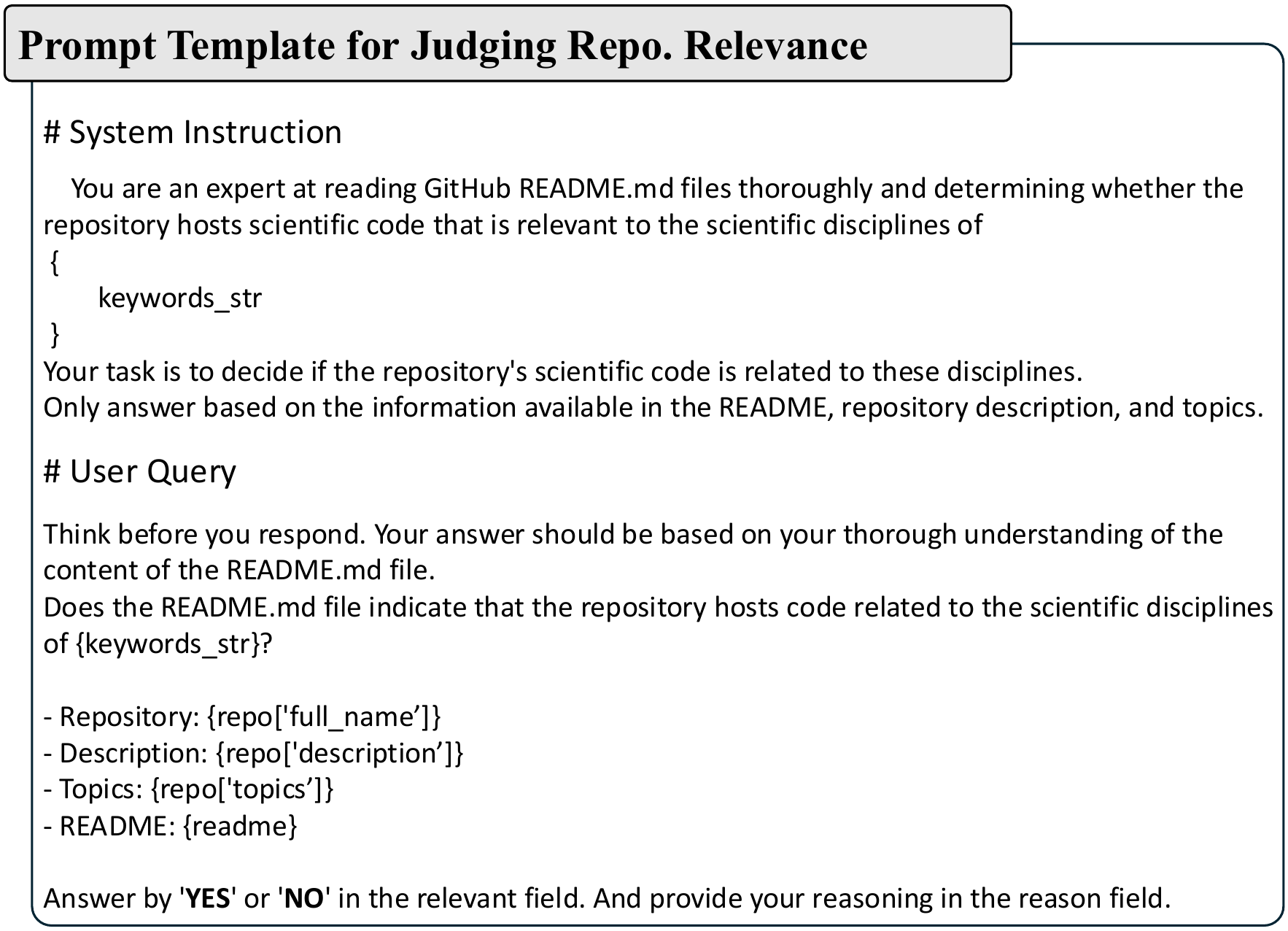}
    \caption{Prompt Template for Judging Repo. Relevance.}
    \label{fig:prompt2}
\end{figure*}

\begin{figure*}[t]
    \centering
    \includegraphics[width=\linewidth]{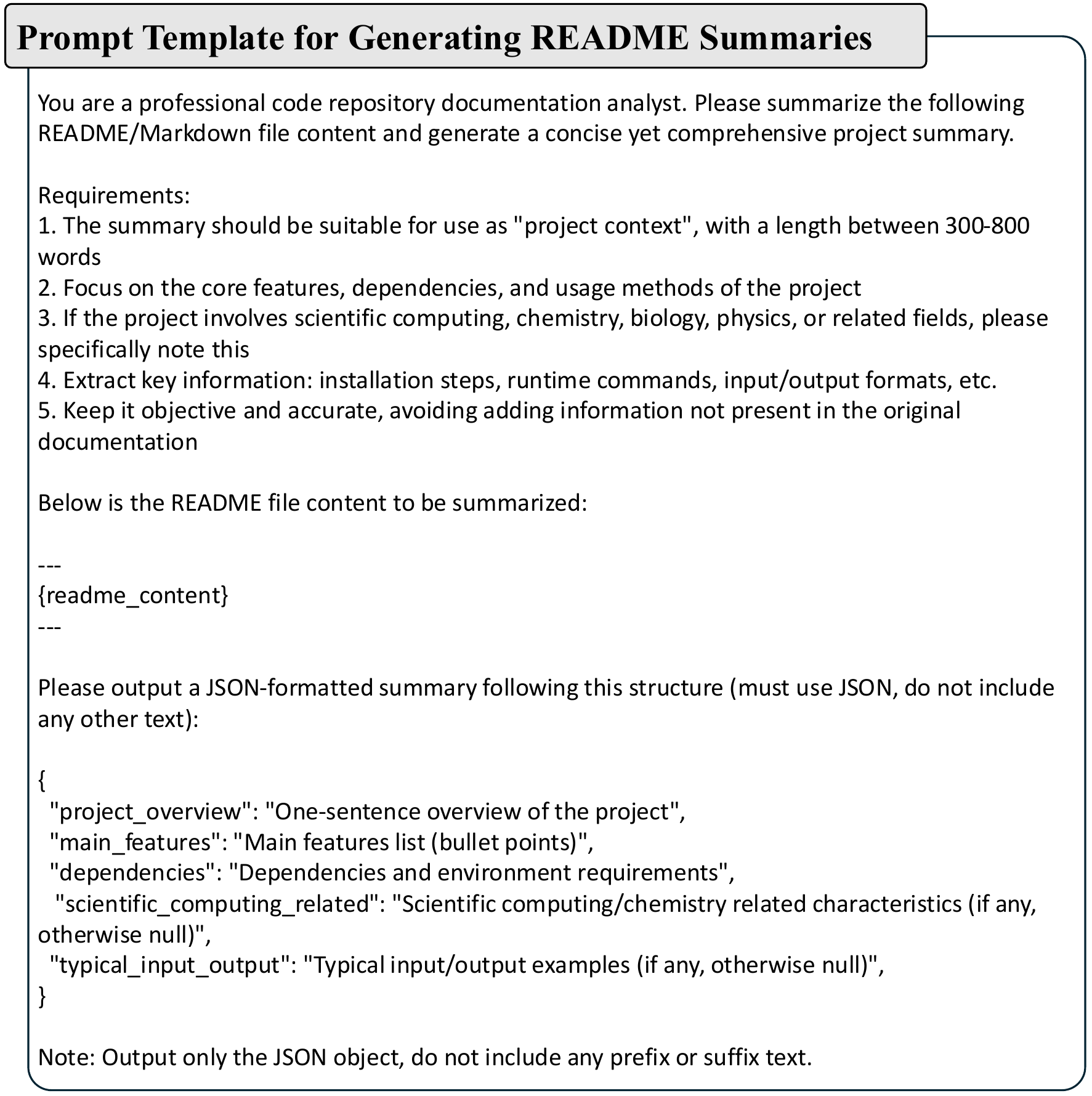}
    \caption{Prompt Template for Generating \texttt{README} Summaries.}
    \label{fig:prompt3}
\end{figure*}

\begin{figure*}[t]
    \centering
    \includegraphics[width=\linewidth]{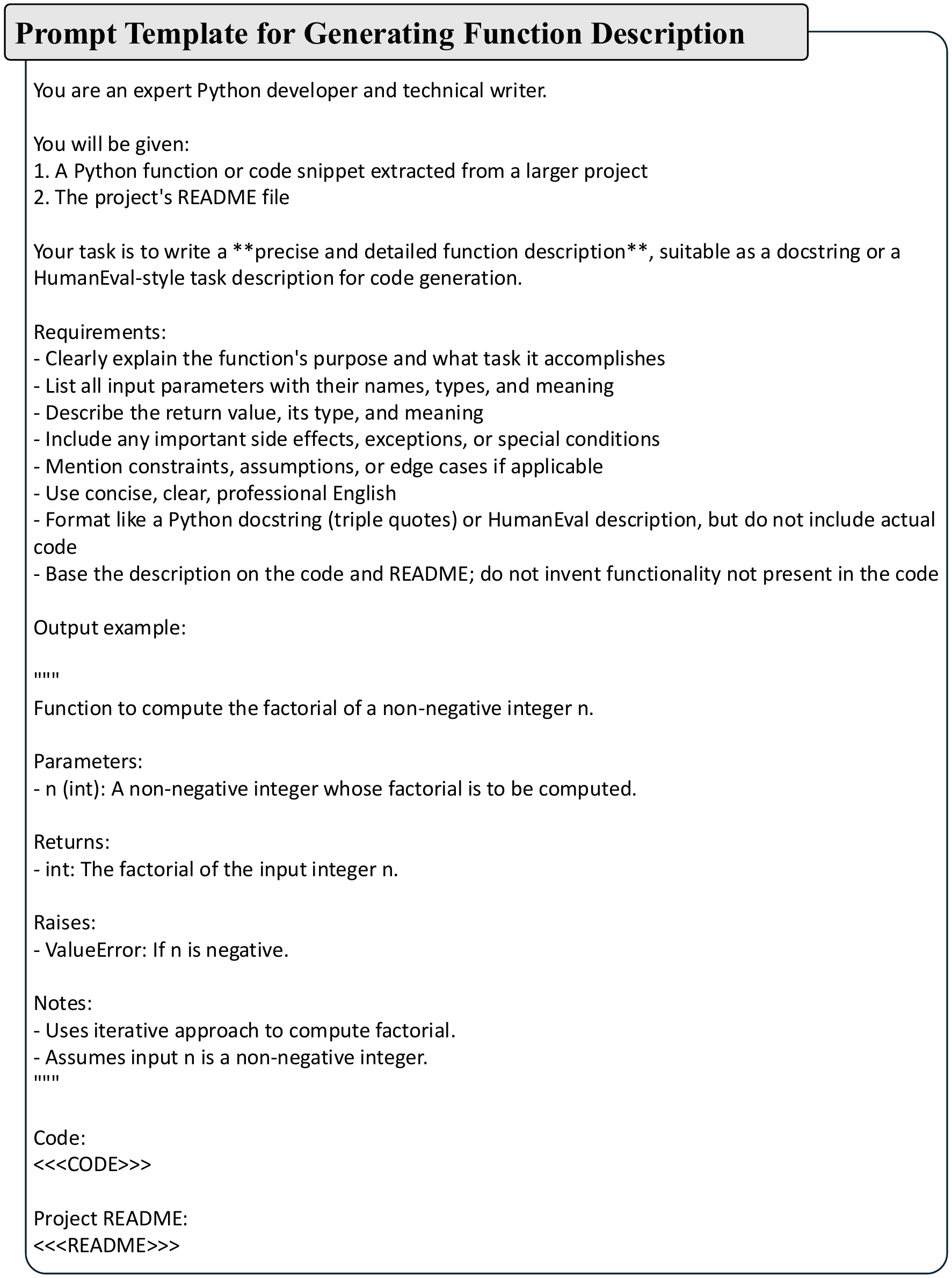}
    \caption{Prompt Template for Generating Function Description.}
    \label{fig:prompt4}
\end{figure*}

\begin{figure*}[t]
    \centering
    \includegraphics[width=\linewidth]{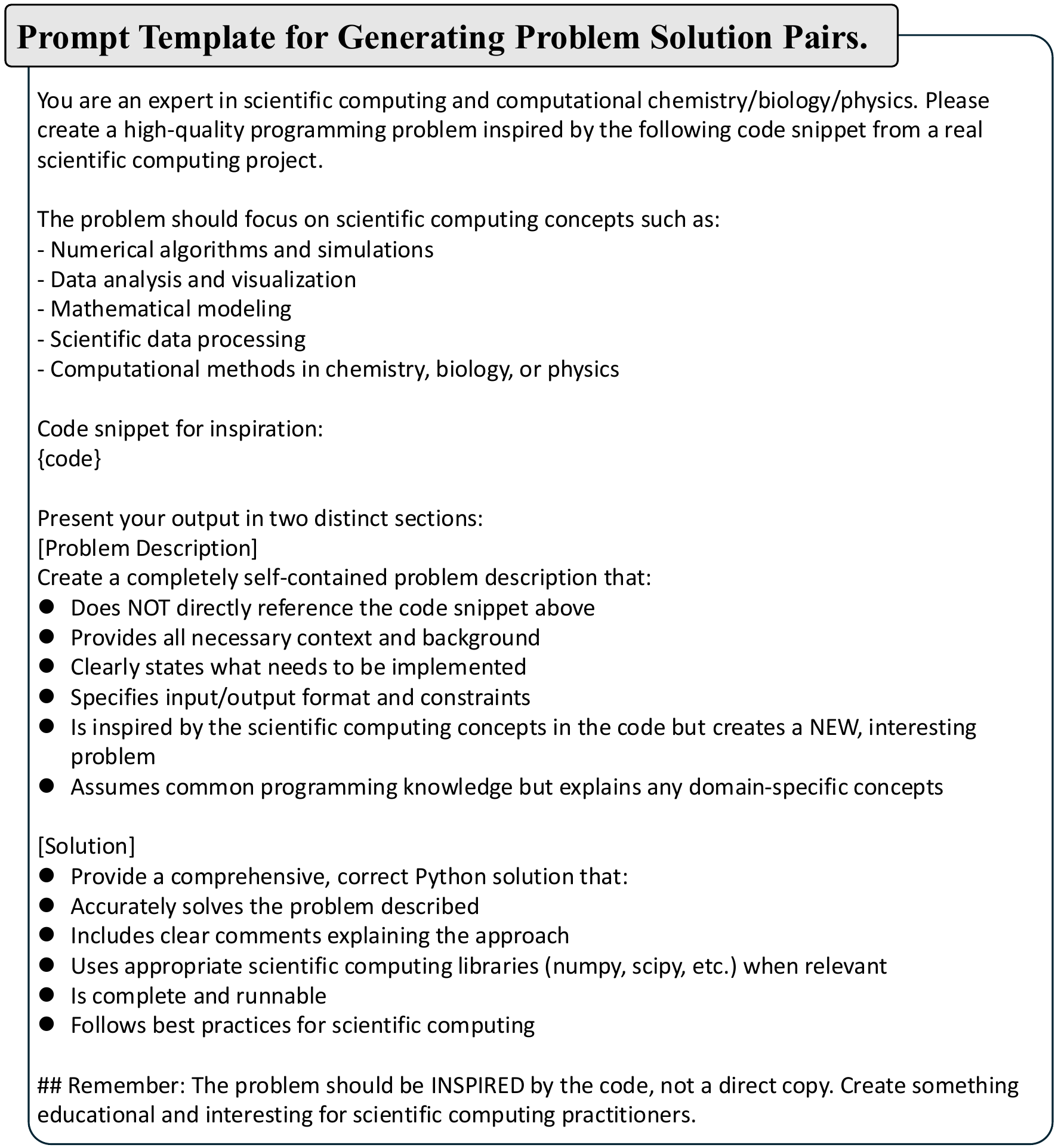}
    \caption{Prompt Template for Generating Problem Solution Pairs.}
    \label{fig:prompt5}
\end{figure*}

\begin{figure*}[t]
    \centering
    \includegraphics[width=\linewidth]{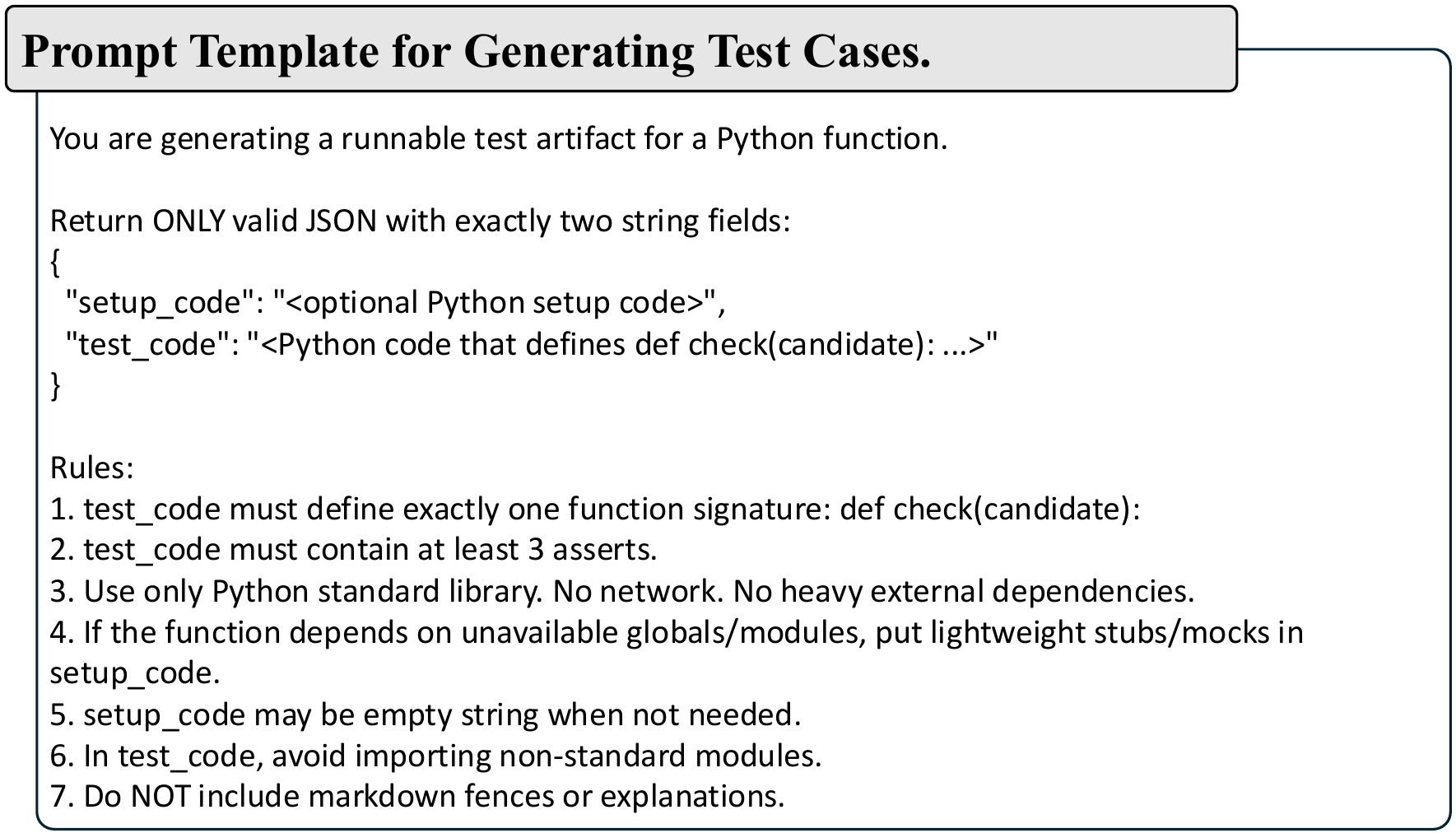}
    \caption{Prompt Template for Generating Test Cases.}
    \label{fig:prompt6}
\end{figure*}

To construct the F1 repository-level dataset, we design two lightweight prompts to assist the identification of domain-relevant scientific repositories. 
Figure~\ref{fig:prompt1} shows the prompt template used for expanding scientific keywords. 
Given a set of initial discipline keywords, the prompt asks the model to generate a small set of closely related terms. 
These expanded keywords are then used to retrieve candidate repositories from GitHub, enabling broader coverage of domain-specific projects while avoiding overly generic terms.

Figure~\ref{fig:prompt2} presents the prompt used for judging repository relevance. 
The model is instructed to analyze repository metadata, including the \texttt{README} content, description, and topics, and determine whether the repository contains code relevant to the target scientific disciplines. 
The output is a binary decision (``\texttt{YES}'' or ``\texttt{NO}'') together with a brief justification. 
These prompts help filter repositories and ensure that the collected projects are closely aligned with the intended scientific domains.

\subsection{Prompt Template for F2}
\label{sec:append_f2}

We design a prompt to generate structured summaries from repository \texttt{README} files. 
As shown in Figure~\ref{fig:prompt3}, the prompt instructs the model to analyze the \texttt{README}/\texttt{Markdown} content and produce a concise yet comprehensive project summary. 
The generated output follows a predefined \texttt{JSON} schema, including fields such as project overview, main features, dependencies, scientific-computing characteristics, and typical input/output information. 
This structured summary serves as lightweight project context for downstream stages of the dataset construction pipeline.

\subsection{Prompt Template for F3}
\label{sec:appd_f3}
To construct the F3 function-instruction dataset, we design a prompt to generate structured natural-language descriptions for individual functions. 
As illustrated in Figure~\ref{fig:prompt4}, the prompt provides the extracted function code together with the corresponding project \texttt{README}, and instructs the model to produce a precise docstring-style description. 
The generated description summarizes the function’s purpose, input parameters, return values, and potential constraints or edge cases, while remaining grounded in the provided code and documentation. 
This process pairs real functions with evidence-constrained instructions, forming the core data used in the F3 dataset.

\subsection{Prompt Template for F4}
\label{sec:appd_f4}
To construct the F4 problem–solution dataset, we design a prompt that generates scientific programming tasks inspired by real code snippets. 
As illustrated in Figure~\ref{fig:prompt5}, the prompt provides a code fragment extracted from the repository and instructs the model to create a self-contained problem description together with a corresponding Python solution. 
The generated problems focus on scientific computing concepts such as numerical algorithms, data analysis, or computational modeling, while avoiding direct references to the original code. 
Therefore, each instance forms a complete problem–solution pair that reflects the underlying scientific computing ideas present in the source repository.

\subsection{Prompt Design for Executable Benchmark Construction}
\label{sec:appd_f5}

To construct the executable benchmark, we design a prompt that generates runnable test artifacts for Python functions. 
As illustrated in Figure~\ref{fig:prompt6}, the prompt instructs the model to produce a \texttt{JSON} object containing optional setup code and a test function that validates the candidate implementation. 
The generated test code must define a single \texttt{check(candidate)} function and include multiple assertions while relying only on the Python standard library. 
These constraints ensure that the resulting test cases are lightweight, executable, and suitable for automatic evaluation in downstream tasks.

\clearpage
%%%%%%%%%%%%%%%%%%%%%%%%%%%%%%%%%%%%%%%%%%%%%%%%%%%%%%%%%%%%

\end{document}